\documentclass[twocolumn]{aastex631}

\usepackage{hyperref}
\usepackage{textcase}

\hypersetup{linkcolor=cyan,citecolor=cyan,filecolor=cyan,urlcolor=cyan}
\usepackage{enumitem}
\shorttitle{Hot Rocks Survey IV: LTT 3780~\texorpdfstring{\MakeLowercase{b}}}
\shortauthors{Allen et al.}

\begin{document}

\title{Hot Rocks Survey IV: Emission from LTT 3780~b is consistent with a bare rock}

\author[0000-0002-0832-710X]{Natalie H. Allen}\altaffiliation{NSF Graduate Research Fellow}
\affiliation{Department of Physics and Astronomy, Johns Hopkins University, 3400 N. Charles Street, Baltimore, MD 21218, USA}
\correspondingauthor{Natalie H. Allen}
\email{nallen19@jhu.edu}

\author[0000-0001-9513-1449]{N\'estor Espinoza}
\affiliation{Space Telescope Science Institute, 3700 San Martin Drive, Baltimore, MD 21218, USA}
\affiliation{Department of Physics and Astronomy, Johns Hopkins University, 3400 N. Charles Street, Baltimore, MD 21218, USA}

\author[0000-0001-8274-6639]{Hannah Diamond-Lowe}
\affiliation{Space Telescope Science Institute, 3700 San Martin Drive, Baltimore, MD 21218, USA}
\affiliation{Department of Space Research and Space Technology, Technical University of Denmark, Elektrovej 328, 2800 Kgs.\,Lyngby, DK}

\author[0000-0002-6907-4476]{João M. Mendonça}
\affiliation{Department of Space Research and Space Technology, Technical University of Denmark, Elektrovej 328, 2800 Kgs.\,Lyngby, DK}
\affiliation{Department of Physics and Astronomy, University of Southampton, Highfield, Southampton SO17 1BJ, UK}
\affiliation{School of Ocean and Earth Science, University of Southampton, Southampton, SO14 3ZH, UK}

\author[0000-0002-9355-5165]{Brice-Olivier Demory}
\affiliation{Center for Space and Habitability, University of Bern, Gesellschaftsstrasse 6, 3012 Bern, Switzerland}
\affiliation{ARTORG Center for Biomedical Engineering Research, University of Bern, Murtenstrasse 50, CH-3008, Bern, Switzerland}
\affiliation{Space Research and Planetary Sciences, Physics Institute, University of Bern, Gesellschaftsstrasse 6, 3012 Bern, Switzerland}

\author[0000-0003-0854-3002]{Am\'{e}lie Gressier}
\affiliation{Space Telescope Science Institute, 3700 San Martin Drive, Baltimore, MD 21218, USA}

\author[0000-0003-2775-653X]{Jegug Ih}
\affiliation{Space Telescope Science Institute, 3700 San Martin Drive, Baltimore, MD 21218, USA}

\author[0000-0002-8938-9715]{Mark Fortune}
\affiliation{School of Physics, Trinity College Dublin, University of Dublin, Dublin 2, Ireland}

\author[0000-0003-3829-8554]{Prune C. August}
\affiliation{Department of Space Research and Space Technology, Technical University of Denmark, Elektrovej 328, 2800 Kgs.\,Lyngby, DK}

\author[0000-0002-0931-735X]{Måns Holmberg}
\affiliation{Space Telescope Science Institute, 3700 San Martin Drive, Baltimore, MD 21218, USA}

\author[0000-0002-2160-8782]{Erik Meier Valdés}
\affiliation{Department of Physics, University of Oxford, Keble Road, Oxford, OX1 3RH, UK}

\author[0009-0001-6868-6171]{Merlin Zgraggen}
\affiliation{Centre for Space and Habitability, University of Bern, Gesellschaftsstrasse 6, 3012 Bern, Switzerland}

\author[0000-0003-1605-5666]{Lars A. Buchhave}
\affiliation{Department of Space Research and Space Technology, Technical University of Denmark, Elektrovej 328, 2800 Kgs.\,Lyngby, DK}

\author[0000-0002-6523-9536]{Adam J. Burgasser}
\affiliation{Department of Astronomy \& Astrophysics, UC San Diego, 9500 Gilman Drive, La Jolla, CA 92093, USA}

\author[0000-0003-0652-2902]{Chloe Fisher}
\affiliation{Department of Physics, University of Oxford, Keble Road, Oxford, OX1 3RH, UK}

\author[0000-0002-9308-2353]{Neale P. Gibson}
\affiliation{School of Physics, Trinity College Dublin, University of Dublin, Dublin 2, Ireland}

\author[0000-0003-1907-5910]{Kevin Heng}
\affiliation{Ludwig Maximilian University, Faculty of Physics, Scheinerstr. 1, Munich D-81679, Germany}
\affiliation{ARTORG Center for Biomedical Engineering Research, University of Bern, Murtenstrasse 50, CH-3008, Bern, Switzerland}
\affiliation{University College London, Department of Physics \& Astronomy, Gower St, London, WC1E 6BT, United Kingdom}
\affiliation{University of Warwick, Department of Physics, Astronomy \& Astrophysics Group, Coventry CV4 7AL, United Kingdom}

\author[0000-0001-8981-6759]{Jens Hoeijmakers}
\affiliation{Lund Observatory, Division of Astrophysics, Department of Physics, Lund University, Box 118, 221 00 Lund, Sweden}

\author[0000-0003-4269-3311]{Daniel Kitzmann}
\affiliation{Space Research and Planetary Sciences, Physics Institute, University of Bern, Gesellschaftsstrasse 6, 3012 Bern, Switzerland}

\author[0000-0001-7216-4846]{Bibiana Prinoth}
\affiliation{Lund Observatory, Division of Astrophysics, Department of Physics, Lund University, Box 118, 221 00 Lund, Sweden}

\author[0000-0002-4227-4953]{Alexander D. Rathcke}
\affiliation{Department of Space Research and Space Technology, Technical University of Denmark, Elektrovej 328, 2800 Kgs.\,Lyngby, DK}

\author[0000-0003-2528-3409]{Brett M. Morris}
\affiliation{Space Telescope Science Institute, 3700 San Martin Drive, Baltimore, MD 21218, USA}



\begin{abstract}
It is an open question whether small planets around M dwarfs are able to maintain atmospheres. The Hot Rocks Survey aims to address this question by observing 9 rocky exoplanets orbiting M dwarfs with MIRI emission photometry to constrain the onset of atmospheres. In this paper, we present two MIRI F1500W (15$\mu$m) eclipses of LTT 3780~b, an ultra-short period super-Earth ($P=0.768$ d, $R=1.325 \,R_\oplus$, $M = 2.46\,M_\oplus$) that receives 111x Earth's instellation, the highest in the survey. We find a combined eclipse depth of $312\pm38$ ppm, which is consistent between different data reduction and analysis assumptions, bolstering our confidence in the eclipse detection. This eclipse depth is consistent with the thermal emission from a bare rock surface, with a dayside temperature of $T_d=1143^{+104}_{-99}$ K, $98\pm9$\% of the maximum temperature predicted for a zero albedo, zero heat redistribution blackbody. We are able to confidently rule out CO$_2$-based atmospheres down to 0.01 bar surface pressure to greater than 3$\sigma$ (ruling out an approximately Mars-like atmosphere). We are unable to rule out a pure H$_2$O 1 bar atmosphere, though we argue that this composition is unlikely on such a highly irradiated planet, nor O$_2$ atmospheres due to the lack of features in the bandpass, though we can put constraints on CO$_2$-mixture atmospheres. As a potential bare rock, we consider a variety of surface composition models, but are unable to distinguish between them. However, LTT 3780~b is an excellent target for follow-up \textit{JWST} observations to determine its surface composition and rule out additional atmospheric compositions.
\end{abstract}



\vspace{2mm}
\section{Introduction} \label{sec:intro}
Over the past few decades, we have discovered the atmospheric composition of dozens of exoplanets. However, despite numerous efforts, definitive evidence for an atmosphere around a terrestrial exoplanet ($R \lesssim 1.6 R_\oplus$) has never been detected. Early efforts to observe terrestrial exoplanets in transmission with \textit{JWST} have mostly yielded either flat spectra with no detectable features, which are consistent with large mean molecular weight atmospheres, high cloud decks, or no atmospheres at all \citep[e.g.,][]{lustig-yaeger_2023, may_2023, kirk_2024}, or slopes/features that are consistent with the presence of inhomogeneities on the surfaces of the host stars and therefore require further study to confirm the presence of any atmospheric signals \citep[e.g.,][]{moran_2023, lim_2023, Radica_2024}. There are a few cases for which an atmosphere provides the tentative best fit to the data, and does not seem to be caused by signals from the host star, from \textit{JWST} transmission spectra, but all require more observations to validate \citep[][]{Gressier_2024,Bello-Arufe_2025}.

On the other hand, some of the most conclusive studies thus far on the state of terrestrial exoplanet atmospheres are based on the planet's thermal emission, utilizing phase curves and/or observations during secondary eclipse. Pre-\textit{JWST}, Spitzer observations of LHS 3844~b suggested that the planet's 4-5 $\mu$m photometric phase curve signal and dayside temperature were consistent with a bare rock \citep{kreidberg_2019}. A similar lack of atmosphere was suggested for GJ 1252~b with eclipse observations using the same Spitzer filter \citep{Crossfield_2022}. With \textit{JWST}, the same lack of substantial atmosphere has been suggested using MIRI 15 $\mu$m secondary eclipse measurements of TRAPPIST-1~c, and 12.8 and 15 $\mu$m for TRAPPIST-1~b \citep{greene_2023, zieba_2023, Ducrot_2025}. Observations with MIRI LRS have concluded likewise for a number of other rocky planets \citep[][]{Zhang_2024, Wachiraphan_2024,  Xue_2024,WeinerMansfield_2024}. In contrast, 55 Cnc e has had claims of an atmosphere with JWST from NIRCam and MIRI LRS secondary eclipse observations, but these observations are rife with systematics due to the brightness of the target's host star, and will require additional observations to confirm \citep[][]{Hu_2024, Patel_2024}. However, 55 Cnc e is a member of a planet population with equilibrium temperatures hot enough to form a rock vapor atmosphere over a molten surface ($T_{eq}\gtrsim2000$ K), distinct from the other cooler planets mentioned, which instead would only have an atmosphere if they manage to maintain a volatile envelope.

The conclusions that these rocky planets either lack or maintain significant atmospheres mainly relies on the simple assumption that the presence of an atmosphere will allow heat recirculation from the permanent dayside to the permanent nightside. Observationally, this corresponds to a smaller eclipse depth, as the dayside temperature of the planet would be cooler with an atmosphere than it would be as a bare rock incapable of circulating heat away from the substellar point \citep[see e.g.,][]{Mansfield_2019, Koll_2022}. Additionally, the Spitzer 4.5 $\mu$m and \textit{JWST} 15 $\mu$m imaging filter fall on strong CO$_2$ bands, which will make the eclipse especially shallow in those wavelengths and therefore increase our detection sensitivity to CO$_2$. As CO$_2$ is predicted as a common component of rocky secondary atmospheres from formation and evolution models \citep[see e.g.,][]{Herbort_2020} and has such a strong opacity even at relatively low abundances, it is an especially promising avenue for the detection of terrestrial atmospheres.

The success of this technique motivated the Hot Rocks Survey (\textit{JWST} GO 3730, PI H. Diamond-Lowe, Co-PI J. M. Mendonça). This program is observing 9 terrestrial exoplanets orbiting early- to mid-M dwarfs with the 15 $\mu$m MIRI imaging filter in secondary eclipse, across a variety of insolation levels, to search for the presence of atmospheric recirculation, and if found, determine which parameters correlate with the onset of an atmosphere (see \citealp{August_2025}, \citealp{Meier_Valdes_2025}, Fortune et al. submitted). In this work, we present the observations of LTT 3780~b, an ultra-short period \citep[$P = 0.768$ d, defined as an orbital period less than one day, e.g.,][]{Sahu_2006} super-Earth ($R=1.325 \,R_\oplus$, $M = 2.46\,M_\oplus$) orbiting a M4 star \citep{Bonfanti_2024}. Also known as TOI-732~b, with an equilibrium temperature of 903 K and a corresponding insolation 111x that of Earth's \citep{Bonfanti_2024}, it is the most highly irradiated target in this survey. As the inner planet to the temperate sub-Neptune LTT 3780~c on the other side of the radius valley (which is being observed by \textit{JWST} in transmission by GO 3557, PI N. Madhusudhan), this system holds potential for understanding the formation and evolution of a diverse multi-planet M dwarf system.

This paper is laid out as follows: \autoref{sec:obs} describes our \textit{JWST} MIRI observations, and \autoref{sec:data} our two data reductions of these observations. In \autoref{sec:fitting}, we go over our multitude of eclipse fitting tests, which culminate in a consistent final eclipse depth between reductions and fitting methods. Our discussion follows in \autoref{sec:discussion}, which covers our confidence of the detection of the eclipse in \autoref{sec:detect} and our comparison to both atmospheric and surface composition models in \autoref{sec:models}. We briefly consider future observation potential in \autoref{sec:future} before finishing with our conclusions in \autoref{sec:conclusion}.

\section{Observations}\label{sec:obs}
We observed two back-to-back eclipse events of LTT 3780~b with \textit{JWST} using the MIRI photometric imaging  F1500W (15 $\mu$m) filter: 4 May 2024 (Observation 15 of GO 3730) and 5 May 2024 (Observation 14 of GO 3730). Both observations used 22 groups per integration and 1717 integrations per exposure, for a total exposure time of 11828 s. This encompasses approximately equal time in and out of the eclipse, plus an additional 1.5 hours at the beginning of the observation to account for both the one hour for the \textit{JWST} phase constraint and a half hour to allow for detector settling. Each observation is composed of 5 segments of data. Due to the brightness of the target, we use the 256 subarray to allow for a substantial number of groups without saturating the detector\footnote{As is suggested in the \href{https://jwst-docs.stsci.edu/jwst-mid-infrared-instrument/miri-observing-strategies/miri-tso-recommended-strategies}{\textit{JWST} User Documentation.}}.

\section{Data Reduction}\label{sec:data}
\subsection{\textit{transitspectroscopy} Reduction}\label{sec:reduction}
We use the code \textit{transitspectroscopy} \citep{transitspectroscopy}, which was developed for analyzing transmission spectra for JWST. Despite the name, the code has been adapted to analyze eclipse photometry as well, as presented here. 

\begin{figure}
    \centering
    \includegraphics[width=\linewidth]{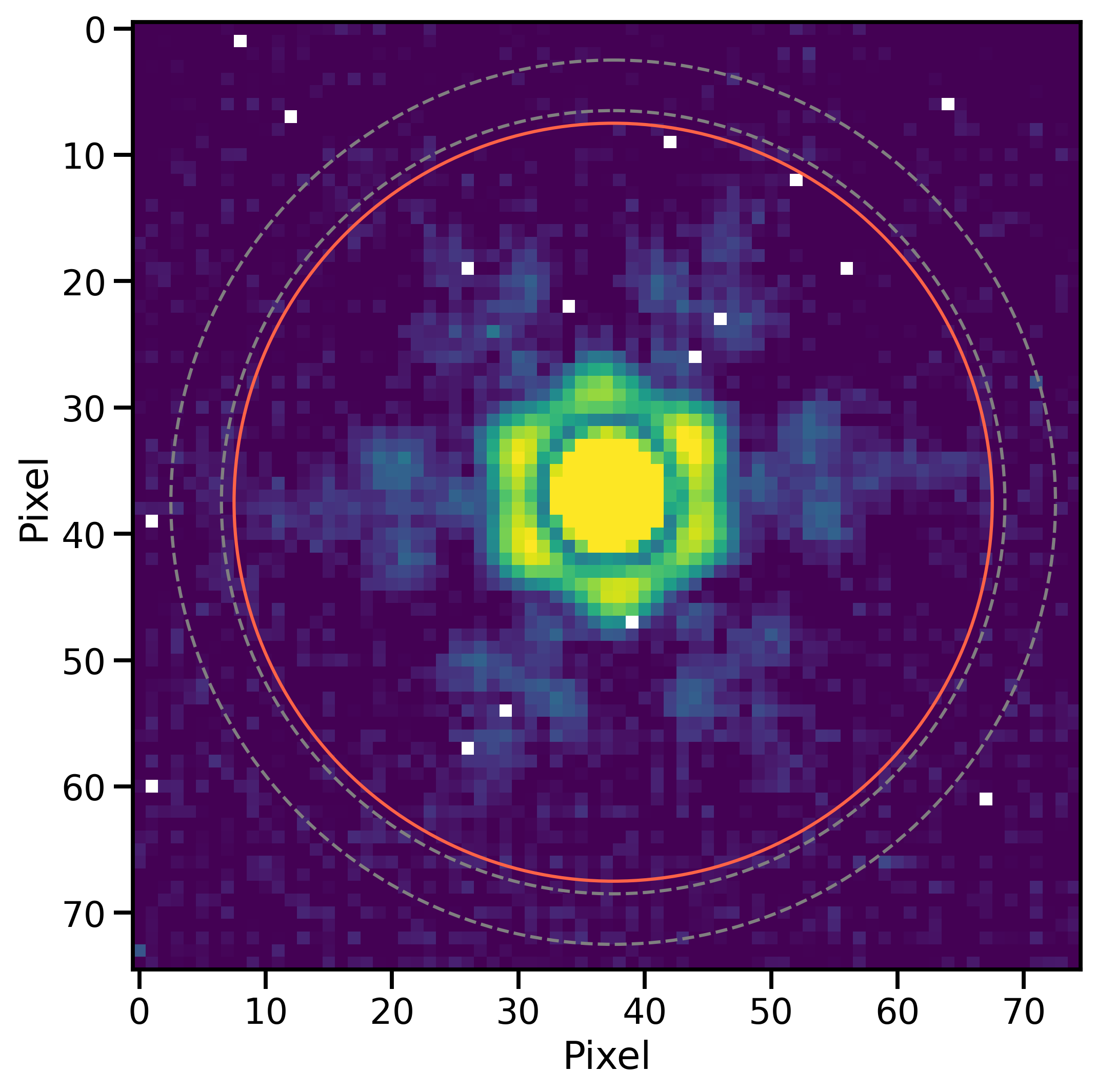}
    \caption{A single MIRI imaging integration of LTT 3780~b, taken from Observation 15, cut from 90 to 165 pixels in each direction of the whole frame. Shown in red is the aperture radius used for optimal extraction (30 pixels), and in grey dashed is the annulus used for background subtraction (31-35 pixels).}
    \label{fig:frame}
\end{figure}

To process the data from the raw \textit{uncal.fits} files into the equivalent of the \textit{rateints.fits}, we mostly use the default \textit{JWST} Calibration pipeline v.1.14.0 \citep{Bushouse_2023} with the Calibration References Data System (CRDS) version 11.17.19 and context \textit{jwst\_1229.pmap}, but with the custom jump step from \textit{transitspectroscopy} \citep{transitspectroscopy}. This custom jump step processes all of the observation's segments collectively rather than individually, which can otherwise sometimes cause offsets between processed segments. A processed integration ready for photometric extraction is shown in \autoref{fig:frame}. 

To begin the photometric extraction, we take a 75x75 pixel box roughly centered on our target star (90 to 165 pixels in both x and y directions), which is the only region we consider in the following steps and for which LTT 3780 is the only star in the box. We find our target centroid position using \textit{DAOStarFinder} from \textit{photutils} \citep{photutil}. We take the box and background-subtract each integration by subtracting the median value from an annulus spanning 31 to 35 pixels around the centroid position (the grey dashed annulus in \autoref{fig:frame}).\footnote{Note that we tested multiple methods of background subtraction, such as using other defined annuli and building a background model through interpolating the background signal across the detector outside of our spectral extraction aperture, but saw a negligible difference in the resulting integrations and light curves.} Using these background subtracted integrations, we build an empirical point-spread function (ePSF) by taking the median of all the integrations in time and then dividing the resulting median integration frame by the median pixel value, such that the pixel values in the final median integration frame (which will act as weights for each pixel in the next step) sum to 1. 

\begin{figure*}
    \centering
    \includegraphics[width=0.8\linewidth]{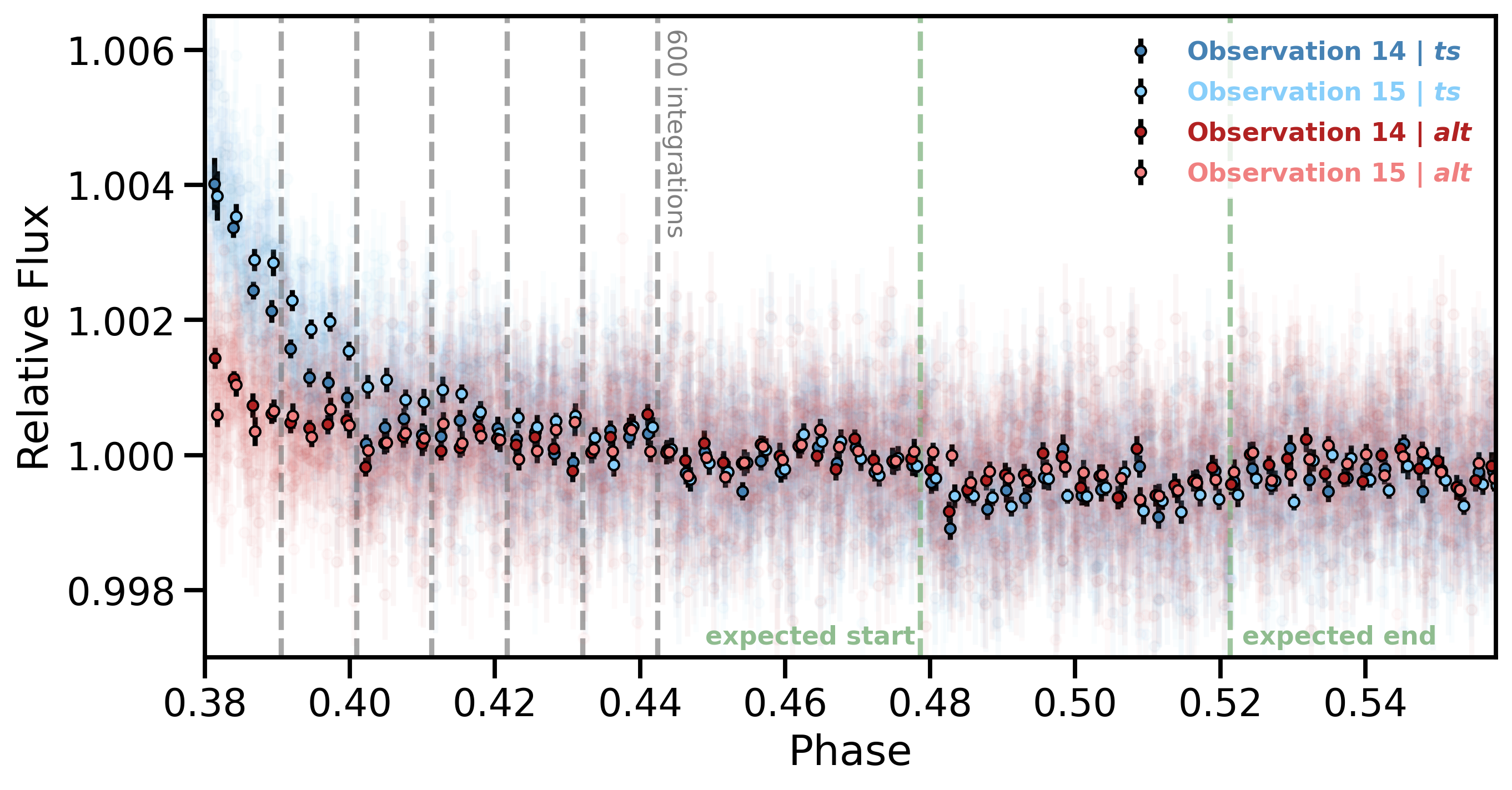}
    \caption{Two MIRI 15 $\mu$m photometric light curves of LTT-3780~b, shown both at native time resolution and binned for visualization, for both the \textit{transitspectroscopy} (ts) and alternate (alt) reductions. Significant deviations from the expected smooth light curve are seen, and are largely in agreement between reductions and between visits, though the initial ramp strength varies between reductions and visits, as discussed in \autoref{sec:compare}. Lines are shown every 100 integrations, to give context to the tests done in \autoref{sec:fitting-cut}. Also shown are the expected eclipse start and end time for a circular orbit.}
    \label{fig:light-curve}
\end{figure*}

Next, we look at the time series of each individual pixel in our box to reject outliers, by median filtering with a window of 21 integrations (approximately 2.5 minutes, though the specific window used does not significantly affect the result) and rejecting any values that differ by this median filter by greater than $5\sigma$, where $\sigma$ is the standard deviation of the residuals between the median filter and the pixel time series. Any rejected value is replaced by the median filter value at the pixel's location in the time series. There are only a small number of rejected outlier integrations (6 for Observation 14, 8 for Observation 15). After removing outliers, we again find the source, apply the background subtraction (as described above), and multiply each integration in the time series with the ePSF, which acts as pixel weights for ``optimal" photometry. We note that the centroid position is incredibly stable throughout our observations, such that this small jitter will not have an effect on the use of the ePSF weights. With these background-subtracted, weighted pixels, we do aperture extraction using a large 30 pixel radius (the red circle in \autoref{fig:frame}). The pixel weights allow us to have this large aperture to capture parts of the PSF beyond the main Airy ring without adding significant additional noise from the background pixels included in this radius, as they will have negligible weight. With this step, we obtain our final photometric light curves, which are shown in \autoref{fig:light-curve}. The function used for this reduction is available in \textit{transitspectroscopy} \citep{transitspectroscopy}.

\subsubsection{\textit{WebbPSF} testing}
We also attempted to use \textit{WebbPSF}\footnote{Newer versions now called STPSF: \url{https://stpsf.readthedocs.io}} \citep{Perrin_2012} to create a model PSF with which to do our PSF extraction, but found that the PSF model resulted in a poor fit to the center of the PSF. The model significantly underpredicted the flux in this region 
and since the center pixels of the PSF contain a large fraction of the overall flux \citep[see e.g.,][]{Libralato_2024}, this resulted in a significant increase in the scatter of the data. We found the flexibility of the ePSF pixel weighting as implemented in our pipeline works better with our observations.

\subsection{Alternate reduction} \label{sec:brice_red}
To test the robustness of our results, we conducted an independent alternate reduction of the data. We performed the detector-level calibration using the \textit{JWST} Calibration pipeline v.1.13.4 \citep{Bushouse_2023}, with CRDS version 11.17.22 and context \textit{jwst\_1263.pmap}. We ran the recently-implemented emicorr step to subtract the 10.04 Hz EMI noise present in MIRI Imaging data. All subsequent steps (saturation detection, linearity model, dark current) were conducted in a standard fashion. Before proceeding with the ramp fitting, we employed the jump detection step as implemented in \textit{transitspectroscopy} v.0.4.0 \citep{transitspectroscopy}. We elected to discard the first and last groups during ramp fitting which often display flux inconsistencies with all-group ramp fit model (each integration has 22 groups in total). We completed the data reduction with \textit{JWST} Calibration pipeline's stage 2 resulting in high-level calibrated data files (calints) that we use as inputs for the next steps.

We first identified bad pixel values that we replaced with values interpolated using flux values from nearest neighbors with a Gaussian weighting. We then computed the star centroid position and performed simple aperture photometry in a set of apertures ranging from 3 to 20 pixels in radii, with a final radius of 5 pixels decided through a comparison of red noise contribution at the light curve fitting stage (see \autoref{sec:brice_fit}). We used a background aperture located at 30 pixels from the source with an annulus of 15 pixels to avoid secondary contribution from the source's PSF. With this method, we obtain our final light curve, seen in \autoref{fig:light-curve}.

\section{Eclipse Fitting}\label{sec:fitting}
\subsection{transitspectroscopy Fits}\label{sec:ts_fit}
We carry out our eclipse fits using \textit{juliet} \citep{juliet}, and our sampling using the \textit{dynesty} dynamic nested sampler \citep{dynesty}. All \textit{juliet} parameters are given as named in the code for clarity. We fix all of our system parameters besides the eclipse depth (which is given a normal prior of [0, 10,000] ppm) to those found in \citet{Bonfanti_2024}. We use the time of central transit and calculate the associated light travel time to fit for the eclipse timing rather than fitting for time of central eclipse. \textit{juliet} also has three instrumental parameters, \textit{mdilution} (dilution factor, fixed to 1.0), \textit{mflux} (offset in relative flux, normal distribution [0.0, 0.1]), and \textit{sigma\_w} (additional jitter, log-uniform [0.1, 1000.0]), which is defined for each observation. All of these parameters are in terms of relative flux, as the light curves are normalized relative to their median value before fitting. Next, we test different systematics models.


\subsubsection{Systematics model testing}\label{sec:fitting-systematics}
We first look for any parameters that may correlate with the systematics seen in the data. We test the periodicity of the centroid position (x and y shift), the \textit{roundness1}, \textit{roundness2} and \textit{sharpness} parameters (different measures of the shape of the PSF) obtained from source finding with the \textit{DAOStarFinder} function from \textit{photutils} \citep{photutil}, and the guide star flux and FWHM \citep[analyzed with \textit{spelunker};][]{spelunker} with our observational data using a Lomb-Scargle Periodogram, but could not identify any matching frequencies. Given no obvious external decorrelation parameter was found, we elected to use time as our regressor and test the use of linear, exponential, and Gaussian Process (GP) systematic models \citep[first presented for transiting exoplanet atmosphere analysis in][and used broadly in the field since]{Gibson_2012}, though we always include the exponential model since we know there is an underlying systematic signal that is approximately exponential in nature \citep[see e.g.,][]{Bouwman_2023}. We use the \textit{celerite} GP kernels \citep{celerite}, and try the exponential, approximate Matern $3/2$, and Matern multiplied by exponential kernels.

The additive linear model has a single associated parameter for the regressor weight per parameter (\textit{theta}) per observation. The exponential is of the form $A\,\,e^{-t/\tau}$, where $t$ is the time from the beginning of the observation, and the $A$ and $\tau$ terms are fit for (additional additive exponential terms were tested, but never preferred by the data). Both are given wide uniform priors of $(-1,1)$. All GP kernels share the same first parameter (\textit{GP\_sigma}), which is the GP amplitude. The exponential and Matern kernels then have parameters \textit{GP\_timescale} and \textit{GP\_rho}, respectively, which are the time or length-scale of the GP, while the Matern multiplied by exponential kernel has both of these parameters. We allow a wide prior range for all systematics parameters, defining a separate systematics model for each observation with a uniform prior of $(-100,100)$ on \textit{theta}, and a log-uniform prior of $(10^{-8}, 10^{-2})$ (corresponding to 0.1 to 10,000 ppm) on \textit{GP\_sigma}, and of $(10^{-8}, 10^2)$ on \textit{GP\_timescale} and \textit{GP\_rho}. 

Through these tests, we found that the preferred systematics model is dependent on the number of removed integrations from the beginning of the observation. Therefore, we consider the number of removed integrations from the beginning of the observation as we continue our systematics model testing. 

\subsubsection{Removal of beginning integrations} \label{sec:fitting-cut}
The behavior and method of fitting the strong systematics ramp at the beginning of MIRI observations may have a significant effect on the resulting eclipse depth. To find the optimal amount of removed integrations, and test our ability to model the systematics, we try removing 100, 200, 300, 400, 500, and 600 integrations. For reference, 100 integrations is approximately 11.5 minutes. We also did tests without removing any integrations, but the beginning of the exponential slope is so steep that the systematics models have a difficult time fitting it together with the rest of the integrations, and so we do not consider it further.  For each number of removed integrations, we again carry out the full test of systematics models explained above. We do this for both the \textit{transitspectroscopy} reduction and the alternate reduction. Our full test results are given in \autoref{tab:tests}.

\begin{figure*}
    \centering
    \includegraphics[width=0.49\linewidth]{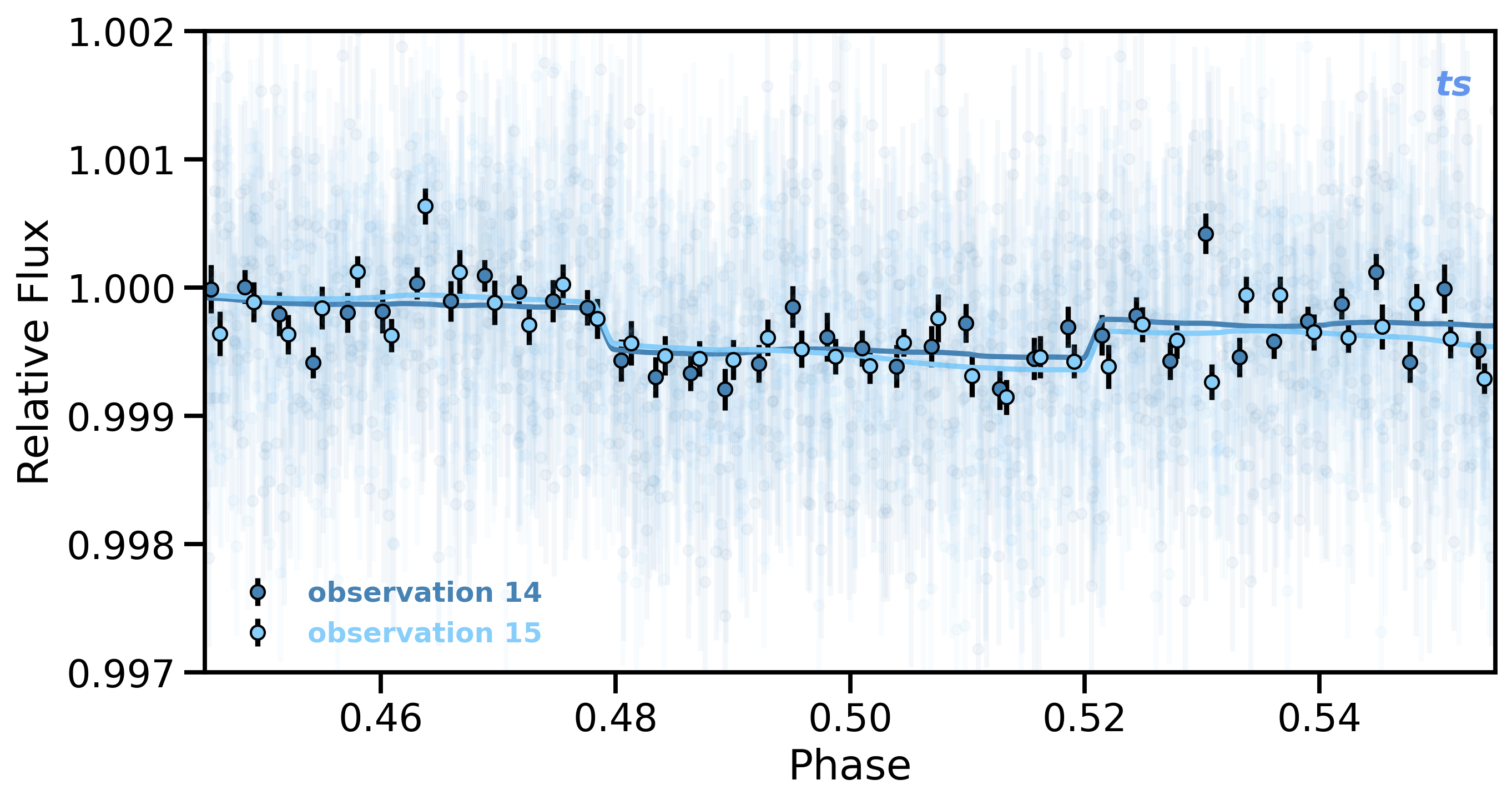}
    \includegraphics[width=0.49\linewidth]{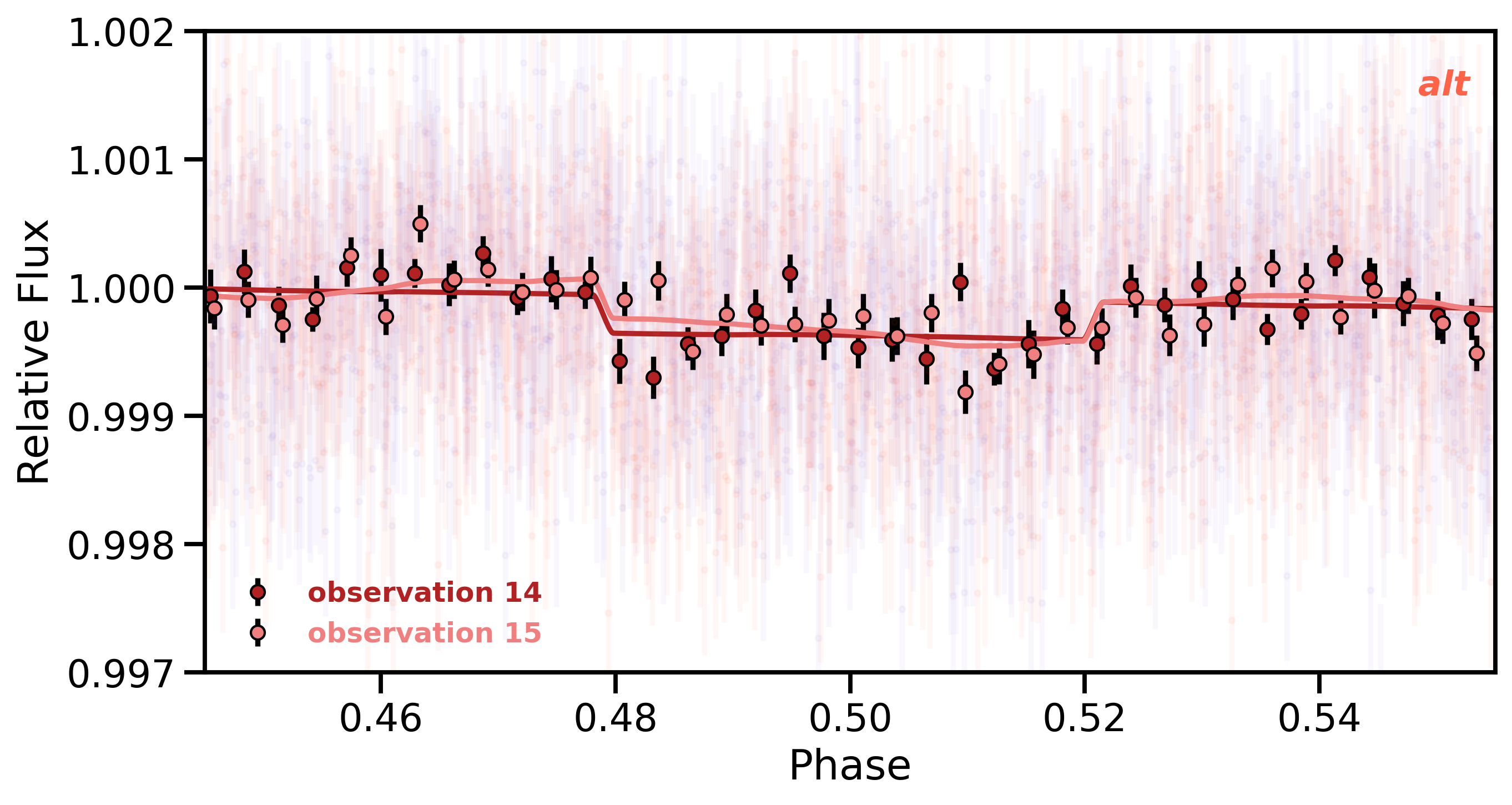}
    \includegraphics[width=0.49\linewidth]{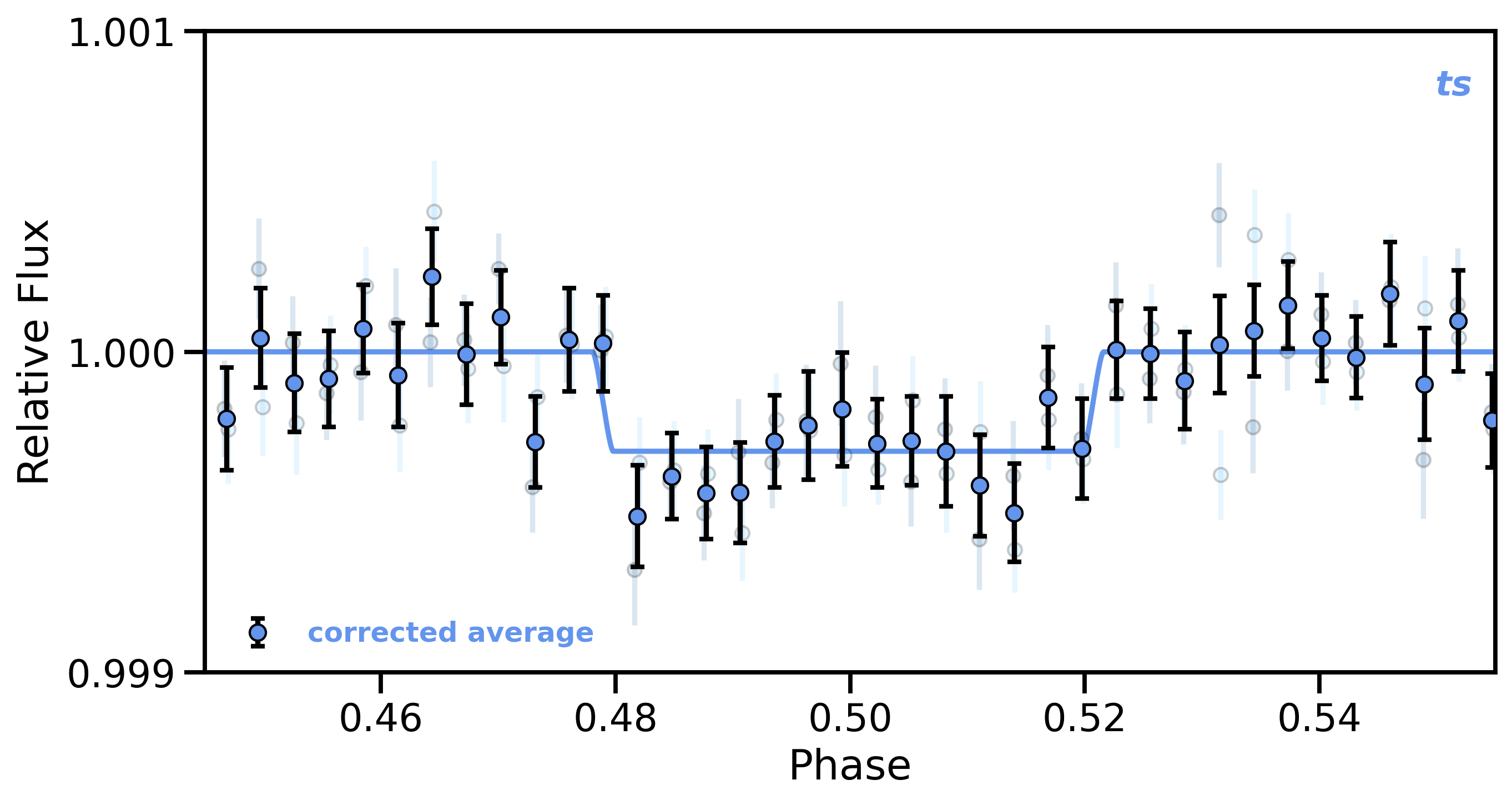}
    \includegraphics[width=0.49\linewidth]{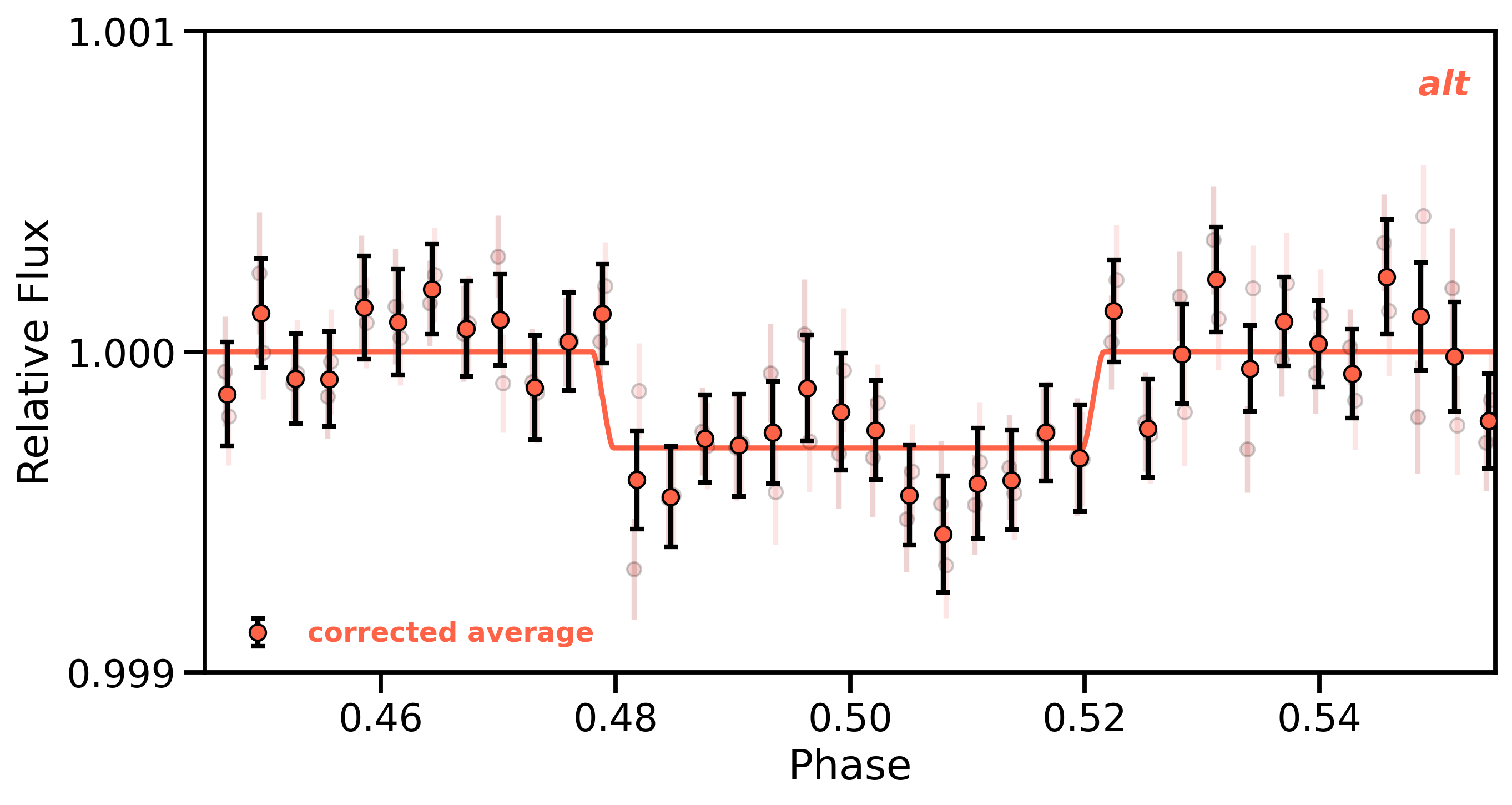}
    \caption{Final joint light curve fit for the two observations of LTT 3780~b, where 600 integrations are removed from the beginning of the observation. Top row: Both observations, with an exponential model and exponential kernel GP used to detrend the systematics. We note that the actual final eclipse depth is calculated using the Bayesian model average between the GP + exponential model and pure exponential model, but we show the more complex model as it encompasses the exponential model as well. Bottom row: Average between the two observations, with systematics model removed.}
    \label{fig:lc-fitting}
\end{figure*}

\begin{deluxetable*}{c l c c| c c}
    \tablecaption{Summary of the light curve fitting tests. GP models all use an exponential kernel, as it was consistently the best-fitting kernel. $\Delta$ ln Z is calculated by taking the difference of the ln Z values output by the \textit{dynesty} nested sampling. \label{tab:tests}}
    \tablehead{\multicolumn{2}{c}{} & \multicolumn{2}{c}{\textit{transitspectroscopy}} & \multicolumn{2}{c}{alternate reduction} \\
    \colhead{Removed integrations} &
    \colhead{Systematics model} & \colhead{Joint eclipse depth (ppm) } & \colhead{$\Delta$ ln Z} & \colhead{Joint eclipse depth (ppm) } & \colhead{$\Delta$ ln Z}}
    \startdata
     100 & exponential & $265\pm 36$ & 2.8 & $313\pm37$ & 0.0\\
     & linear + exponential & $232\pm40$ & 33.7 & $314\pm36$ & 15.9\\
     & GP + exponential  & $301\pm90$ & 0.0 & $323\pm55$ & 0.9\\
     & linear + GP + exponential & $283\pm89$ & 14.8 & $317\pm65$ & 20.4\\
     \hline
     200 & exponential & $272\pm37$  & 8.2 & $318\pm35$ & 0.0 \\
     & linear + exponential & $317\pm38$  & 45.6 & $285\pm32$ & 13.8 \\
     & GP + exponential  & $295\pm77$ & 0.0 & $326\pm55$ & 6.5\\
     & linear + GP + exponential & $270\pm77$ & 16.9 & $307\pm66$ & 24.5\\
     \hline
     300 & exponential & $273\pm39$  & 8.1 & $319\pm35$ & 0.0\\
     & linear + exponential & $230\pm39$ & 32.8 & $297\pm33$ & 15.1\\
     & GP + exponential  & $314\pm65$ & 0.0 & $325\pm55$ & 4.5\\
     & linear + GP + exponential & $298\pm83$ & 15.1 & $311\pm55$ & 21.9\\
     \hline    
     400 & exponential & $343\pm37$  & 1.3 & $310\pm37$ & 0.0\\
     & linear + exponential & $356\pm36$  & 19.3 & $316\pm34$ & 15.4\\
     & GP + exponential  & $301\pm67$ & 0.0 & $310\pm52$ & 3.0\\
     & linear + GP + exponential & $302\pm76$ & 15.9 & $320\pm60$ & 22.9\\
     \hline
     500 & exponential & $339\pm35$  & 1.0 & $315\pm35$ & 0.0 \\
     & linear + exponential & $339\pm35$ & 10.2 & $217\pm37$ & 34.9\\
     & GP + exponential  & $318\pm65$ & 0.0 & $329\pm68$ & 5.0\\
     & linear + GP + exponential & $299\pm70$ & 18.1 & $350\pm53$ & 22.9\\
     \hline
     600 & exponential & $312\pm36$  & 0.0 & $295\pm36$ & 0.0\\
     & linear + exponential & $265\pm39$  & 19.4 & $311\pm36$ & 21.0\\
     & GP + exponential  & $312\pm54$ & 3.3 & $296\pm47$ & 4.0\\
     & linear + GP + exponential & $307\pm58$ & 22.4 & $306\pm76$ & 22.6\\
     \hline
    \enddata

\end{deluxetable*}

\subsubsection{Excluding first and last groups}
In some observations from MIRI, there seem to be detector effects that cause the first group to behave abnormally \citep[see e.g., the RSCD effect, ][]{Morrison_2023}. To test if this is the case in our observations, we calculate new rateints that exclude the first and last group in the calculations. By eye, the removal of the first and last group in the fit has an insignificant effect on the final rateints. However, to test if this has any effect on the fitting process itself, we carry out the same systematics method test as above (in case the removal of these groups changed the underlying systematics in the data, resulting in a different preferred systematics model), but we find the results to be fully consistent with each other across systematics models and ramp fitting methods. Therefore, we continue to use all of the groups for the analysis moving forward.

\subsubsection{Final Joint Fit}\label{sec:fit_final}
We find that no matter the combination of systematics parameters and included integrations, we recover the same eclipse depth to within $1\sigma$ of $\sim 300$ ppm except for a few cases consistent instead at $2\sigma$ but all of which are strongly rejected by the Bayesian evidence ($\Delta$ ln Z $\gtrsim20$). However, the specific configuration does have a strong effect on the calculated uncertainty, as well as the preferred systematics model. In summary, for the \textit{transitspectroscopy} reduction, if cutting 600 integrations, a simple exponential model is slightly preferred by the Bayesian evidence, but if including any more, a GP model is preferred in addition to the exponential, though this is not a strong preference until only 300 integrations are removed. On the other hand, the alternate reduction always prefers the simpler exponential systematics model. This difference is discussed further in \autoref{sec:compare}. Since the two reductions agree most closely when 600 integrations are removed, we choose to report these values as our final best-fit solution to the eclipse depth. To ensure that we are correctly representing the uncertainty associated with the multiple solutions to the eclipse depth, we perform Bayesian model averaging for our final value by weighing each eclipse depth measurement by the relative odds represented by its $\Delta$ ln Z \citep[see e.g.,][]{Hinne_2020}. This effect is small since the only exponential model is quite strongly preferred over the next best-fit of the GP + exponential model, but it increases the uncertainty associated with the \textit{transitspectroscopy} reduction slightly. Our final eclipse depths are $312\pm38$ ppm and $295\pm36$ ppm for the \textit{transitspectroscopy} and alternate reduction, respectively, which will be used to compare to models in \autoref{sec:discussion}. 
Our final fits are shown in \autoref{fig:lc-fitting}. Note that while all results presented here are for joint fits of both observations, we also carried out a similar testing campaign on the individual observations and found results fully consistent with the joint fits. The final best-fit values, for the individual observations and joint fit, for both reductions are given in \autoref{tab:final}. While the specifics of Bayesian statistics (values of ln Z) can be sensitive to the priors used in the analysis, we would like to emphasize that regardless of particular of our ``goodness of fit" determination, almost all of our fit values are consistent to within $\sim1\sigma$, and therefore we are confident in our result.

\begin{deluxetable}{c|cc}[]

    \tablecaption{Final best-fit eclipse depths calculated using the method described in \autoref{sec:fit_final}, for the individual observations and the joint fit, for both the \textit{transitspectroscopy} and alternate reductions. All values are given in ppm.\label{tab:final}}
    \tablehead{ & {\textit{transitspectroscopy}} & {alternate reduction} }
    \startdata
    Observation 14& $303\pm52$ & $284\pm53$\\
    Observation 15& $321\pm51$ & $305\pm51$ \\
    \hline
    \hline
    Joint Fit & $312\pm38$ & $295\pm36$
    \enddata
\end{deluxetable}


\subsection{Comparison between reductions}\label{sec:compare}
As can be seen in \autoref{fig:light-curve}, the two analyses we presented have different instrumental systematics slopes seen in the beginning of the visit. This difference seems to be due to the photometric extraction: optimal photometric extraction in the \textit{transitspectroscopy} reduction vs. simple aperture extraction in the alternate reduction. We support this theory by repeating the \textit{transitspectroscopy} reduction identically, but swapping the optimal extraction routine for a simple aperture extraction with a radius of 5 pixels as was done in the alternate reduction, and found that our resulting light curve and light curve fits were indistinguishable from the alternate reduction.

In the optimal extraction case, the \textit{transitspectroscopy} reduction has a significantly steeper exponential, which also seems to last longer into the visit than in the alternate reduction. This seems to be because the high flux pixels in the center of the PSF are also more heavily affected by systematics, which because of the pixel weighting done during the optimal extraction, are more obvious in the final light curve. 
The effect of this difference in systematics can also be seen in the eclipse fit tests. The \textit{transitspectroscopy} reduction finds the smallest errors when 600 integrations are cut off of the beginning, and quite steadily increases in uncertainty as less integrations are cut (besides the simple exponential case). However, the alternate reduction instead mostly holds constant in eclipse depth uncertainty as more integrations are added in. This is because the steeper, longer slope in the \textit{transitspectroscopy} reduction requires a more significant detrending effort to remove, which amplifies the associated uncertainty as the GP amplitude is increased. 

However, once enough integrations are cut from the beginning of the observation to fully remove the initial ramp, the uncertainties agree much more closely. Whether optimal extraction or simple aperture extraction is a better technique to use for MIRI photometry will require further testing, across datasets and observational configurations. The pixel fitting method presented in Fortune et al. (submitted) represents yet another possible method for photometric extraction, with it's own nuances discussed therein. Therefore, we suggest the testing of multiple photometric extraction methods in datasets of interest to corroborate results -- in this dataset, all eclipse fits, regardless of cut integrations, systematic detrending methods, or reduction, agree to within $\sim2\sigma$, more often to within $\sim1\sigma$ (see \autoref{tab:tests}).  

\subsection{Alternate eclipse fit test}\label{sec:brice_fit}
To ensure that our fitting method is not somehow biased, we carry out an additional, fully independent eclipse fit on the alternate reduction presented in \autoref{sec:brice_red}. 

The extracted data all display a strong ramp with a duration of $\sim$40 min (or $\sim$350 integrations) that we discarded. We find that the remaining data exhibit a downward slope that we fit with a linear trend simultaneously to a planet occultation model in an MCMC framework previously described in e.g., \citet[][]{Gillon_2017, Demory_2023}. We analyze all sets of time-series (each with flux extracted using different aperture sizes as described in \autoref{sec:brice_red}) and find that an aperture radius of 5 pixels yield the smallest red noise contribution, determined by looking at the standard deviation of the residuals versus binning in time, which we find negligible. We measure an occultation depth of 272 $\pm$ 46 ppm. Our analysis using all groups during ramp fitting (opposed to discarding the first and last groups) yield similar results with 273 $\pm$ 48 ppm, albeit with a marginally larger level of correlated noise. Both of these results are fully consistent with the results in \autoref{tab:tests}.

\section{Discussion}\label{sec:discussion}
\subsection{Did we definitively detect the eclipse?}\label{sec:detect}

\subsubsection{Eccentric orbit}
Our observation windows and fits thus far have assumed a zero eccentricity orbit. While this is a logical assumption as this planet's tidal circularization timescale is predicted to be on the order of 1-15 Myr \citep{Cloutier_2020, Bonfanti_2024}, and previous observations have found eccentricities consistent with zero \citep{Nowak_2020, Luque_2022}, we want to test this assumption by 
\begin{enumerate}[label=(\alph*)]
\item letting the eccentricity and argument of periastron vary, 
\item letting the eccentricity, argument of periastron, and time of central transit, which defines the time of central eclipse in our model, vary,
\item fitting for all orbital parameters free to within their uncertainties from 
\citet{Luque_2022} (since \citealt{Bonfanti_2024} assumes a circular orbit in their work)
\end{enumerate}
We use the exponential and GP systematics models for these tests to be conservative, but note that we get the same results (albeit with smaller uncertainties on the eclipse depths) when only the exponential systematics model is used. All other parameters are fit as described in \autoref{sec:ts_fit} and \autoref{sec:fit_final}.

To fit for an eccentric orbit, we use the parameterization $\sqrt{e} \,\, cos(\omega)$ and $\sqrt{e} \,\, sin(\omega)$\footnote{which are $secosomega$ and $sesinomega$, respectively, in \textit{juliet}} for eccentricity $e$ and argument of periastron $\omega$. This is essentially to allow uniform priors for $e$ and $\omega$ without nonphysical values given their ranges of definition, such that they aren't biased towards larger values \citep[for more information, see e.g., the discussion in][]{Eastman_2013, juliet}. We let each of these values vary from -1 to 1, which encompasses their entire parameter space, but we require the eccentricity to be 0.3 or less (meaning we reject all samples of $secosomega$ and $sesinomega$ that result in an eccentricity greater than 0.3) as the sampler can get stuck at large eccentricity values, and this is still much higher than other values from the literature. For case a), when fitting for an eccentric orbit, we find an eclipse depth fully consistent with the no eccentricity value ($377\pm83$ and $336\pm74$ for the \textit{transitspectroscopy} and the alternate reduction, respectively), though slightly disfavored with a $\Delta$ ln Z of 2.2 against the circular model. 

For case b), we allow $T_0$ vary up to $\pm 0.0032$ days, which is 10x larger than the published uncertainty on $T_0$ from \citet{Bonfanti_2024}, in addition to allowing the eccentricity to be up to 0.3. We are able to find essentially the same eclipse location, offset by about a minute \citep[note that as the uncertainty on T$_0$ is approximately 30 seconds,][this is within expectations for a circular orbit]{Bonfanti_2024}, and eclipse depths that are fully consistent with our best-fit results, and case a) ($354\pm71$ and $355\pm79$ for the \textit{transitspectroscopy} and the alternate reduction, respectively). 

For case c), we use the priors on all parameters from \citet{Luque_2022}. Our results are once again consistent with the rest of the tests, with eclipse depths of $354\pm65$ ppm and $323\pm58$ ppm for the \textit{transitspectroscopy} and the alternate reduction, respectively, and a similar offset on T$_0$ of about a minute, within expected uncertainties from previous measurements. With these tests all in agreement, we are confident that we are indeed finding the eclipse when assuming zero eccentricity. 

\subsubsection{Systematics-only model}
To test the significance of our detection further, we try to fit the data using only the linear and exponential kernel GP systematics model. This results in a $\Delta ln\,\,Z$ of -5.4 and -7 for the \textit{transitspectroscopy} and alternate reductions, respectively, which corresponds to ``strong" evidence \citep{Trotta:2008}. Therefore, we are confident that we are truly detecting the eclipse.

\subsection{Does LTT 3780~b have an atmosphere?}\label{sec:models}
\begin{figure*}
    \centering
    \includegraphics[width=1\linewidth]{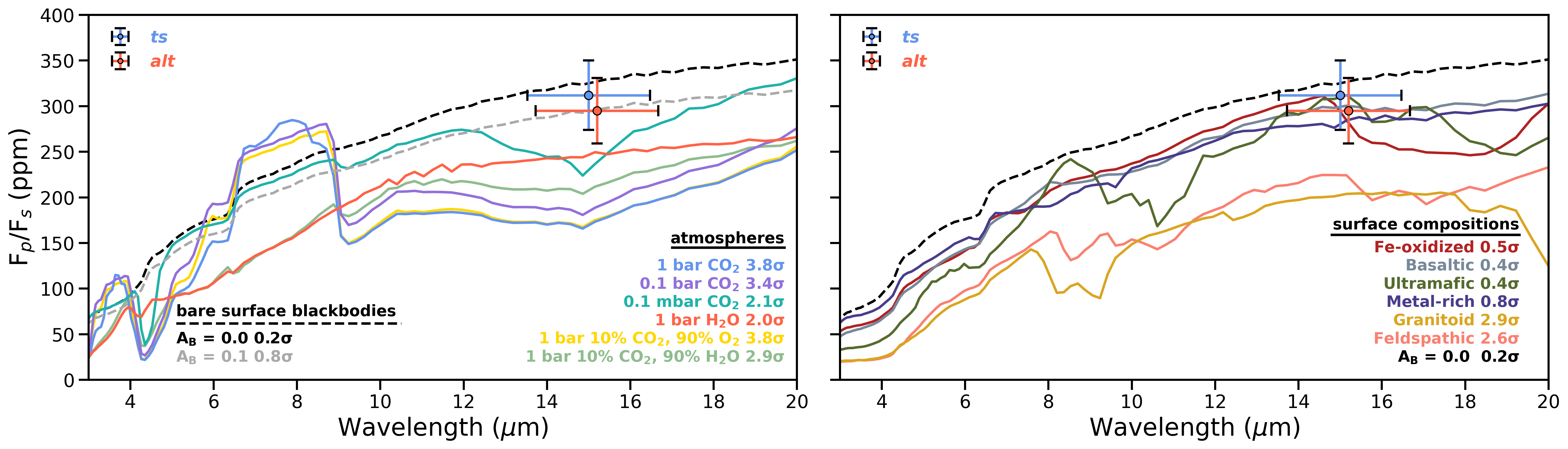}
    \caption{HELIOS model comparison with fit eclipse depth. Note that the alternate reduction has been slightly offset in wavelength for visual clarity. Each model has listed an associated $\sigma$ value for the \textit{transitspectroscopy} reduction. Left: Atmospheric (solid line) and bare surface (dashed line) models. LTT 3780~b is fully consistent with bare surfaces with $A_B$ of 0 and 0.1. We are able to confidently (3$\sigma$) rule out CO$_2$-dominated atmospheres to 0.01 bar, which is approximately consistent with the surface pressure of Mars. We are unable to confidently rule out H$_2$O atmospheres, though we argue these are a-priori unlikely on a highly irradiated target like this, and we are able to rule out O$_2$-CO$_2$ mixture atmospheres to the same total atmospheric surface pressures as the pure CO$_2$ atmospheres due to the lack of O$_2$ features in this bandpass. Right: Surface composition models. We are unable to differentiate between the different surface compositions with this single photometric point due to the considered compositions' lack of significant features in the bandpass, though the more reflective surfaces lie further from our data. }
    \label{fig:model}
\end{figure*}

\subsubsection{Atmospheric Grid Comparison}
To determine the photometric emission consistent with a bare rock with no overlying atmosphere, we estimate the dayside temperature of the planet with
\begin{equation}\label{eq:t_d}
    T_{day} = T_\star\sqrt{\frac{R_\star}{a}}(1-A_B)^{1/4}f^{1/4}
\end{equation}
where $T_\star$, $R_\star$, $a$, $A_B$, and $f$ are the stellar effective temperature, stellar radius, semi-major axis, Bond albedo, and heat redistribution factor, respectively. For the case of no atmosphere, $f$ is $2/3$, corresponding to the flux weighted average temperature of the of the hemisphere with no heat redistribution \citep{Hansen_2008}. We consider bare rock Bond albedo values from 0 to 0.2 (represented as blackbodies), as more reflective surfaces are unlikely on an planet with no atmosphere due to space weathering \citep[see e.g.,][]{Domingue_2014, Brunetto_2015, Mansfield_2019}. 

For atmospheric scenarios, we use models made with \textit{HELIOS} \citep{Malik_2017, Malik_2019a, Malik_2019b, Whittaker_2022} to compare with our photometric emission measurement. \textit{HELIOS} is a 1D plane-parallel model which calculates the planet's temperature structure through the balance between radiative and convective processes and heat redistribution, and then determines the resulting synthetic emission and reflectance spectra. For heat redistribution in the case of an atmosphere, we use the analytical equation from \citet{Koll_2022}, which approximates the $f$ factor as set by the longwave optical depth at the surface, the surface pressure, and the equilibrium temperature of the planet. All atmospheres are calculated with an underlying planetary surface albedo of 0.1. We calculate our stellar models using the BT-Settl (CIFIST) grid \citep{Allard_2011, Allard_2012}, linearly interpolated for our exact stellar parameters. 

Note that we flux calibrate our observations following the methods described in \citet{Gordon_2025} and Fortune et al., submitted, and find a stellar flux of $14.16\pm0.07$ mJy (consistent between the two visits), which is in agreement with the prediction from the BT-SETTL models of $13.77\pm1.08$ mJy, and thus we find that these stellar models are appropriate for use in our atmospheric modeling efforts. However, we do note that it is still possible that the models may disagree with the stellar flux at shorter wavelengths, which could have an impact in the models, but which we are unable to test without further observations at other wavelengths to the same data precision.

We mainly consider atmospheres composed of CO$_2$ since it is so abundant in our own Solar System rocky planets (without oceans), and is also predicted to be a dominant species due to outgassing on planets across a broad range of surface pressures \citep[see e.g.,][and references therein]{Gaillard_2014, Herbort_2020}. This was the motivation to specifically choose observations with the F1500W MIRI filter, which is centered on the strong  15 $\mu$m CO$_2$ feature. In addition, we consider atmospheres composed of mixtures of H$_2$O, another predicted common atmospheric species \citep{Herbort_2020}, though one more susceptible to atmospheric erosion. We also consider atmospheres of O$_2$, which may build up on M dwarf planets after the photodissociation of H$_2$O and subsequent H escape \citep[e.g.,][]{Luger_2015,Schaefer_2016, Bolmont_2017}. All models are cloud-free. We show all of our atmospheric models and Bond albedo-defined bare surface blackbody models in \autoref{fig:model}.


The measured thermal emission at 15 $\mu$m for LTT 3780~b is fully consistent with a bare rock with no atmosphere, though our measurement is unable to differentiate between the different modeled surface Bond albedos of 0 and 0.1. We are able to rule out the CO$_2$-dominated atmospheres at greater than 3$\sigma$ down to surface pressures of 0.01 bar (which is close to the Martian surface pressure of 0.006-0.007 bar), with the 0.1 millibar case shown in \autoref{fig:model} consistent at 2.1$\sigma$ shown as an extreme example (60x-70x thinner than Mars' atmosphere). On the other hand, we cannot rule out H$_2$O atmospheric cases due to the species' lack of strong features in this bandpass, with the 1 bar case consistent at 1.7$\sigma$. However, as stated previously, this species is significantly less likely to remain on a highly irradiated planet than CO$_2$, since C is much more stable to evaporation and loss than H after photodissociation \citep[see e.g.,][]{catling_2007}. O$_2$-CO$_2$ mixture atmospheres are able to be ruled out to the same surface pressures as the pure CO$_2$ atmospheres due to O$_2$'s lack of opacity in this bandpass (see \autoref{fig:model}, where the 1 bar O$_2$-CO$_2$ mixture model overlaps perfectly with the 1 bar CO$_2$ atmosphere over the observed bandpass), tested down to 10\% CO$_2$. This does mean we are unable to rule out pure-O$_2$ composition atmospheres though.  

We can test how robust the atmospheric compositions we are unable to rule out from our observations are through an estimate of the predicted atmospheric mass lost due to photoevaporation. While the saturation phase length, which defines the most XUV-active portion of the star's life, isn't well-known for M dwarfs, recent estimates place the timescale in the hundreds of millions of years for stars like LTT 3780 \citep[M3.5-M4, e.g.,][]{Johnstone_2021, Engle_2024}. Therefore, to be conservative, we assume a saturation timescale of $t_{sat} = 100$ Myr. We estimate the XUV instellation over the saturated phase, $I_{XUV} = 7.1\times10^{20}$ erg/cm$^2$, using equation 6 from \citet{Pass_2025}.

Following the conservative energy-limited escape mass loss estimate from \citet{Hu_2023} for a solar composition atmosphere, the total mass loss over the saturated phase is $1.4\times10^{23}$ kg of atmosphere. This is more than 250x Venus' atmospheric mass \citep[$4.8\times10^{20}$ kg,][]{Gillmann_2024}, lost in 100 million years, much shorter than the estimated age of LTT 3780 \citep[$3.10^{+6.20}_{-0.98}$ Gyr,][]{Bonfanti_2024}. We can estimate the atmospheric mass of LTT 3780~b as $4\pi R_p^2P_{surf}/g_{surf} = 6.536\times10^{18}$ kg ($P_{surf}$/1 bar) \citep{Crossfield_2022}, meaning over 20,000 bars of atmospheric mass could be stripped during the host star's saturated phase. This would erode any primary atmosphere, as well as any outgassing that occurs in the first 100 million years \citep[heavy elements can be carried away by the rapid flow of lighter gas like H, ][]{catling_2007}.

This ignores additional outgassing of atmospheric mass from the planet's interior after the saturated phase, the specifics of which are beyond the scope of this work, but which is predicted to last for billions of years \citep[see e.g.,][]{Dorn_2018, Moore_2020}. Therefore, we also attempt to estimate the current potential for atmospheric mass loss, considering the modern XUV flux from the star. The stars most similar to LTT 3780 from the MUSCLES survey (GJ 436 and GJ 176) both have measured XUV luminosities on the order of $10^{27}$ ergs/s \citep{France_2016}. Using the same equation from \citet{Hu_2023} and a conservative XUV luminosity of $1\times10^{27}$ ergs/s, this is a mass loss rate of $4.3\times10^{11}$ kg/year, or about a bar of atmospheric mass lost every 15 million years for energy-limited hydrodynamic mass loss. This calculation makes many assumptions about things we do not understand well about M dwarfs and secondary atmospheres, such as the XUV flux from other similar M dwarfs being the same as from LTT 3780 and the conditions for energy-limited hydrodynamic atmospheric escape holding, and ignores important considerations like the planetary magnetic field strength or stellar wind interaction. Whether even this substantial mass loss rate would be enough to fully strip LTT 3780 b on its atmosphere also depends on how rapid the outgassing rate is, as well as the initial volatile content of the planet. Even for the most well-studied terrestrial exoplanet system, TRAPPIST-1 \citep[which has a very similar XUV luminosity as GJ 436 and GJ 176,]{Wheatley_2017, Wilson_2021}, there are disagreements in the literature on whether atmospheres on the planets would survive or not given the specific details of the atmospheric mass loss simulations \citep[see e.g.,][though both agree that TRAPPIST-1~b, which has the same orbital separation as LTT 3780~b, is likely to be airless]{Krissansen-Totton_2024, VanLooveren_2024}. There is clearly a large potential for atmospheric mass loss from LTT 3780~b, but whether this would leave the planet completely airless requires more sophisticated mass loss simulations of super Earths, beyond the scope of this work. Therefore, we conclude that while it may be that LTT 3780~b is an airless body as its emission is consistent with a low albedo blackbody, we cannot fully rule out very thin atmospheres nor thick atmospheres for species that do not have opacity in the MIRI F1500W bandpass.

\subsubsection{Surface Composition Comparison}
As our data suggest that LTT 3780~b may be an airless body, we have the opportunity to potentially characterize the bare planetary surface. We run surface models covering a variety of potential compositions with HELIOS, based on \citet{Whittaker_2022} and \citet{Ih_2023}: Fe-oxidized, basaltic, ultramafic, metal-rich, granitoid, and feldspathic. These are shown on the right side of \autoref{fig:model}, also shown with the 0 albedo blackbody for reference. All considered compositions are fully consistent with the observations, as their differences from the 0 albedo blackbody differs by only tens to $\sim$ 100 ppm at 15 $\mu$m. The more reflective granitoid and feldspathic compositions are the furthest from our data, as expected from the high best-fit dayside temperature, and comment that these surfaces are unlikely to form on a bare exoplanet regardless \citep[e.g.,][]{Mansfield_2019}. We note that surface composition models can be more complex and degenerate than those we present here \citep{Paragas_2025}, but as our data quality is insufficient to differentiate even these simple models, we do not explore any more complicated compositions.

\subsubsection{Dayside Temperature}\label{sec:dayside}

We calculate the dayside temperature described by our eclipse depth through  
\begin{equation}
    \frac{F_p}{F_s} = \left(\frac{R_p}{R_s}\right)^2 \frac{\int \frac{\pi B_p(T_{day}, \lambda) M(\lambda)}{hc/\lambda} d\lambda}{\int \frac{F_\star(T_\star, \lambda) M(\lambda)}{hc/\lambda} d\lambda}
\end{equation}
where $B_p$ is the planetary blackbody function, $M$ is the MIRI F1500W throughput \citep[calculated with \textit{pandeia} version 4.0,][]{Pontoppidan_2016}, and $F_\star$ is the interpolated stellar spectrum, made using BT-SETTL (CIFIST) models \citep{Allard_2011, Allard_2012}. Using \textit{dynesty} \citep{dynesty}, we find $T_d = 1143^{+104}_{-99}$ K. If we compare this to the maximum theoretical temperature for LTT 3780~b of 1164 K, corresponding to zero heat redistribution and zero albedo and equivalent to the temperature scaling factor $\mathcal{R}$ from \citet{Xue_2024} and \citet{WeinerMansfield_2024}, we find that LTT 3780~b radiates at $\mathcal{R}=98\pm9$\% of the maximum temperature at 15 $\mu$m. 

\begin{figure}
    \centering
    \includegraphics[width=\linewidth]{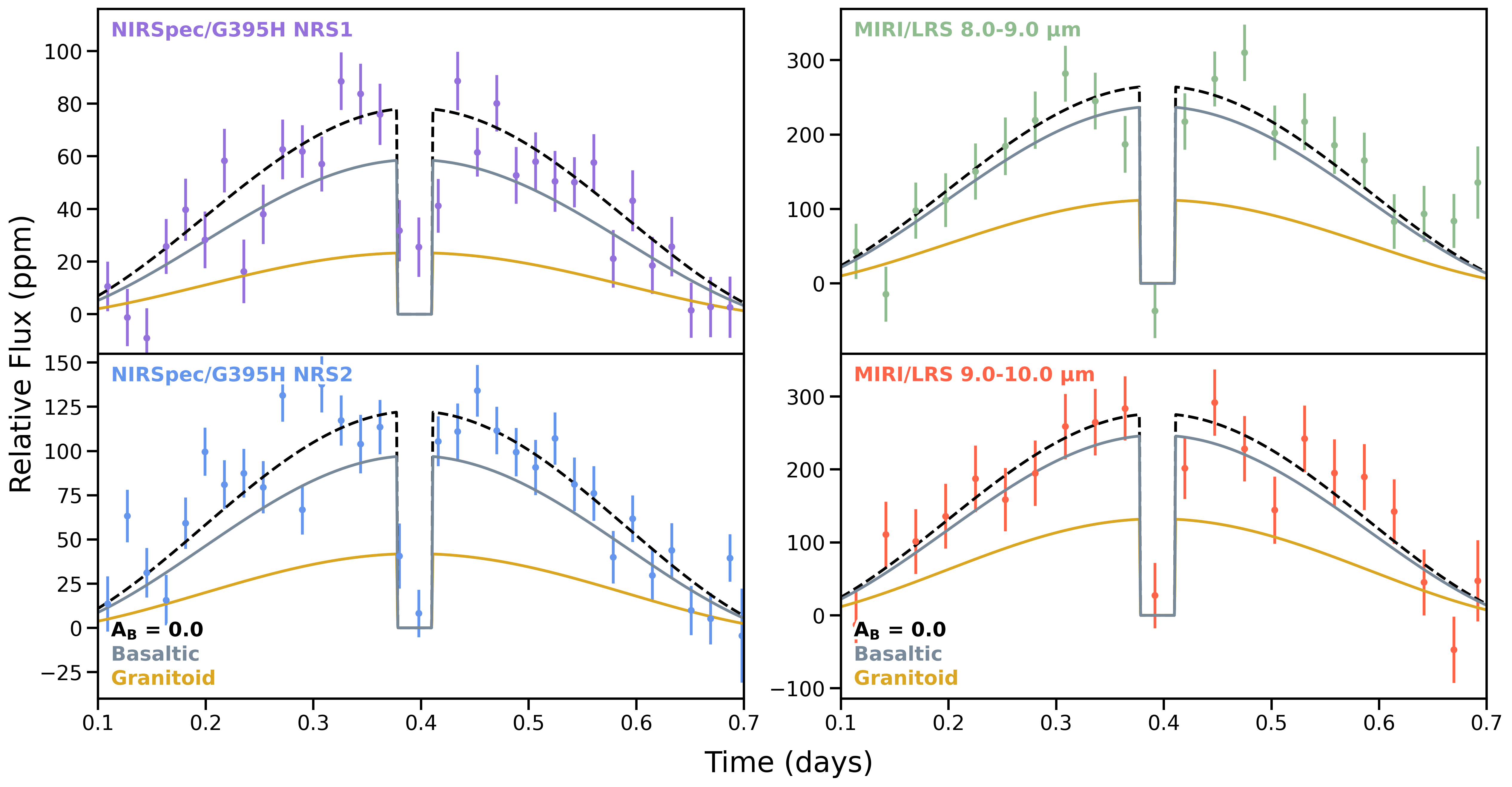}
    \includegraphics[width=0.9\linewidth]{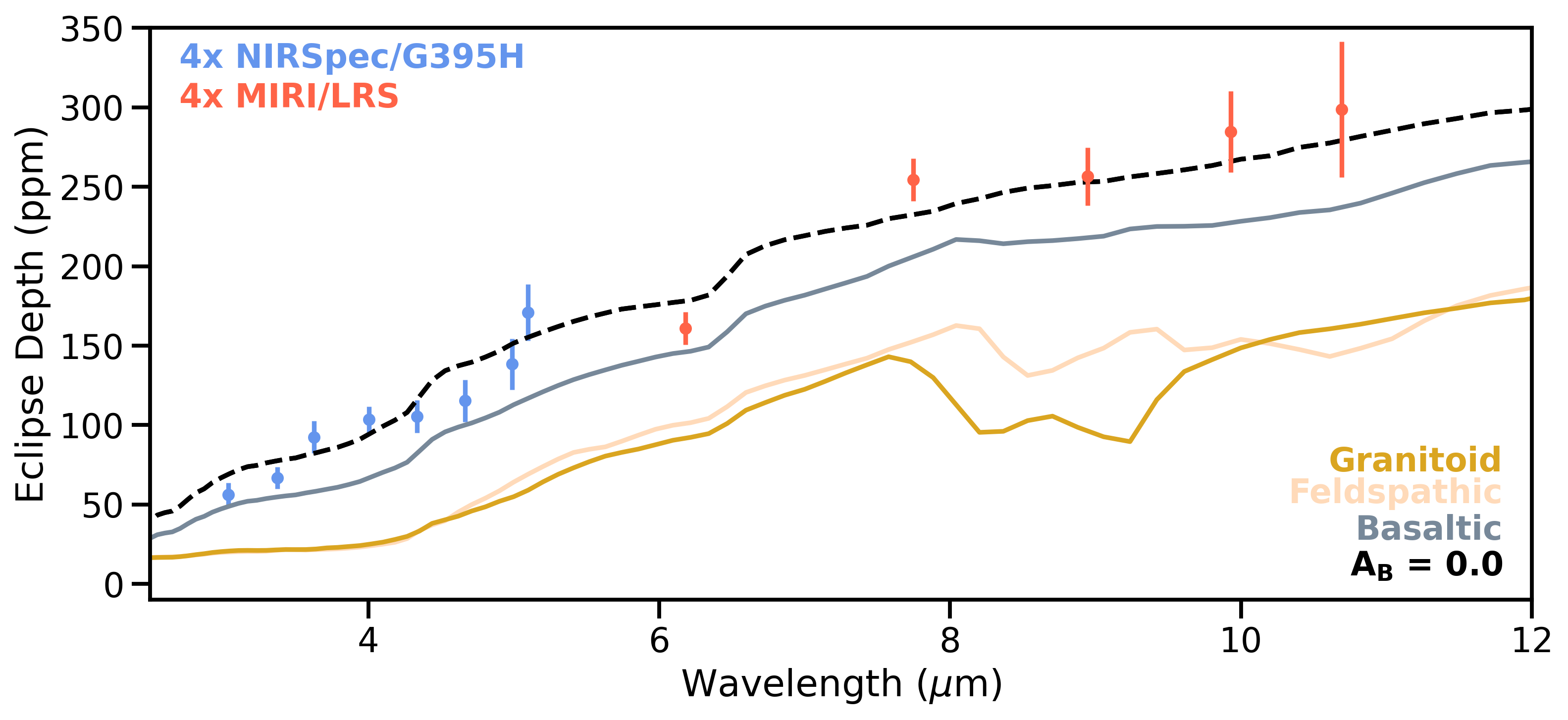}
    \caption{Simulated \textit{JWST} observations to potentially distinguish the surface composition of LTT 3780~b. Top: NIRSpec/G395H (left) and MIRI/LRS (right) phase curves are able to distinguish between a granitoid and zero albedo blackbody surface. Two 1$\mu$m bins are shown for MIRI/LRS, corresponding to the largest signal due to the presence of a broad feature in the granitoid emission (see bottom panel). Bottom: Four NIRSpec/G395H or MIRI/LRS eclipses are also able to distinguish between a granitoid and zero albedo blackbody surface with the same time commitment as a phase curve, but the NIRSpec/G395H observations are also able to differentiate from a basaltic composition.}
    \label{fig:future}
\end{figure}

\section{Future Observation Potential}\label{sec:future}
As an ultra-short period planet ($P=0.768$ d), LTT 3780~b is a promising target for future phase curve observations. In particular, since LTT 3780~b may not have an atmosphere, further observations in emission would allow us to put potential constraints on its rocky surface composition. It is currently not known what the surfaces of these highly irradiated and space weathered surfaces look like, and \textit{JWST} observations present an opportunity to observationally determine the characteristics of rocky surfaces \citep[see e.g., discussion in][]{Paragas_2025}. However, phase curves require significant time investment -- the time required for one phase curve of LTT 3780~b (19 hour orbital period, plus overheads) could instead be used for four eclipse observations (4.7 hours charged time per eclipse for the eclipses presented in this work).

Thus, we simulate potential phase curve and eclipse observations with \textit{JWST} to test the feasibility of future observations to constrain surface composition. For the phase curves, we simulate the precision per integration using the \textit{JWST} Exposure Time Calculator (ETC) for NIRSpec/G395H and MIRI/LRS, which have been previously used for rocky planet phase curve observations. We simulate our phase curves using \textit{starry} \citep{Luger_2019}. For the expected eclipse depth precision, we simulate observations using PandExo \citep{pandexo}. We show our simulations in \autoref{fig:future}. We find that with a NIRSpec/G395H or MIRI/LRS phase curve, or four eclipses with either NIRSpec/G395H or MIRI/LRS (representing equal time commitment), we can distinguish between a zero-albedo blackbody and a globally uniform granitoid or feldspathic surface composition, which are present on Earth's surface as important and sizeable components of the continental crust \citep[e.g.,][]{Rudnick_2003, Laurent_2020}. However, \citet{Mansfield_2019} argue that these surfaces are unlikely on $\sim$Earth-sized, hot exoplanets like LTT 3780~b, and are able to be ruled out for TRAPPIST-1~b \citep{Ih_2023}. Though, with four NIRSpec/G395H eclipses, we find we are able to distinguish between the zero albedo blackbody and a basaltic-type composition as well, which is formed by volcanism on Earth. This is, however, all under the assumption that there is truly no atmosphere on LHS 3780~b. Phase curve observations are able to put additional constraints on the presence or lack of an atmosphere through measurement of thermal emission from the planet's nightside, which can be tied to atmospheric recirculation if significant \citep[see e.g.,][for a non-detection of nightside thermal emission in a rocky planet phase curve]{kreidberg_2019}. Tenuous atmospheres are difficult to distinguish from surface albedos in emission, but we find it unlikely that a thin atmosphere is able to survive on LTT 3780~b considering its high energy history (see discussion in \autoref{sec:models}), and we could detect a substantial O$_2$-based atmosphere using this technique. Whether phase curves or eclipses are better equipped to provide information on rocky exoplanets is under debate in the community \citep[see e.g., discussions in][]{Hammond_2025}.

We note, that these simulations are idealized in that they assume photon noise-limited observations as given by the \textit{JWST} ETC. In reality, early observations with \textit{JWST} utilizing some of these observing modes for observations of eclipses and phase curves of small planets have shown additional systematics, which could be astrophysical or instrumental in nature, that impact the noise budget and therefore the number of observations needed to reach definitive conclusions \citep[see e.g.,][]{Luque_2024, Wachiraphan_2024}. Further publications of observations utilizing these methods, especially the JWST observations specifically targeting LHS 3844~b to do the type of surface composition characterization suggested here with both MIRI LRS and NIRSpec/G395H in phase curve and eclipse (GO 1846, PI L. Kreidberg; GO 4008, PI S. Zieba; GO 7953, PI K. Paragas) will help illuminate the best practices for these types of observations in the future.

\section{Conclusion}\label{sec:conclusion}
We present here two eclipses of the highly irradiated (111x Earth instellation) ultra-short period super-Earth LTT 3780~b observed with the MIRI F1500W imaging mode. We present two separate analyses with different data reduction pipelines, with the most significant difference due to the use of optimal photometric extraction in one and simple aperture photometry in the other, which show significantly different systematic exponential slopes at the beginning of the observation, but agree well in the rest of the lightcurve. In both reductions, we see evidence of systematics, though the systematic exponential slope is different between visits. 

We fit the light curves using a variety of systematics models and varying the number of integrations removed from the beginning of the observation, and find that the preferred systematics model differs depending on the choice of included integrations, and that the corresponding uncertainty depends on both as well. We also include a test of a completely separate light curve fitting procedure to test our methods. However, both reductions and the majority of light curve fits agree on an eclipse depth $\sim$300 ppm. For our final ``best fit" case, we choose to cut 600 integrations and use Bayesian model averaging to take into account the difference in uncertainty found by the GP and non-GP systematics detrending methods, resulting in a final value of $312\pm38$ ppm and $295\pm36$ ppm for the \textit{transitspectroscopy} and alternate reduction, respectively. 

We compare these eclipse depths to a variety of models and find that they are consistent with a bare rock with an albedo of 0.0-0.1 with no atmosphere, corresponding to a dayside temperature of $T_d = 1143^{+104}_{-99}$ K, which is $98\pm9$\% of the maximum theoretical dayside temperature. We are able to rule out a pure CO$_2$ or O$_2$-CO$_2$ mixture atmosphere to 0.01 bar of surface pressure (tested down to 10\% CO$_2$) at 3$\sigma$, though we are unable to rule out a 1 bar H$_2$O atmosphere confidently. However, since H$_2$O is likely to be dissociated on these highly irradiated targets, and we are able to more confidently rule out O$_2$-CO$_2$ than H$_2$O-CO$_2$ atmospheres, we find it unlikely that an atmosphere that was originally composed of a significant fraction of H$_2$O remains. We predict atmospheric mass loss during the saturated regime of the host star's life, finding that over 20,000 bars of atmospheric mass could be lost in just 100 million years, and even with current stellar flux levels, there is the potential for bars of atmospheric mass lost on tens of Myr timescales. However, these mass loss estimates are highly dependent on a variety of planetary and stellar conditions that we do not understand well enough to be confident in. We are therefore unable to put confident constraints on thin atmospheres, nor on pure O$_2$ atmospheres, due to their lack of opacity over the MIRI F1500W bandpass (tested up to a 1 bar surface pressure). 


The emission from LTT 3780~b is consistent with that from a bare rock, emitting at close to the maximum possible temperature. This is not unexpected, given the target's instellation, which is by far the highest in the Hot Rocks Survey. Therefore, LTT 3780~b seems to join a population of hot rocky planets orbiting M dwarfs without significant atmospheres \citep[e.g.,][]{kreidberg_2019, Crossfield_2022, greene_2023, zieba_2023, Zhang_2024, Xue_2024, WeinerMansfield_2024, Wachiraphan_2024}, and once again points towards more temperate planets as our best chance for identifying the onset of rocky planetary atmospheres around these small stars. However, LTT 3780~b represents an excellent opportunity to study the surface composition of a highly irradiated rocky planet through future \textit{JWST} observations, which would also be able to distinguish if a thicker atmosphere of an optically inactive species like O$_2$ was somehow able to remain despite the high energy environment.

\begin{acknowledgments}
We thank our reviewer Nicolas Cowan for a thoughtful review process. This work makes use of observations made with the NASA/ESA/CSA James Webb Space Telescope. The data were obtained from MAST at STScI, which is operated by the Association of Universities for Research in Astronomy, Inc., under NASA contract NAS 5-03127 for \textit{JWST}. These observations are associated with program GO 3730. The specific observations analyzed can be accessed via\dataset[DOI:10.17909/rpt8-jk12]{https://doi.org/10.17909/rpt8-jk12}. We acknowledge the input from the following individuals to the GO 3730 proposal: Can Akin, Andrea Guzmán Mesa, Nicholas Borsato, Meng Tian, Mette Baungaard. NHA acknowledges support by the National Science Foundation Graduate Research Fellowship under Grant No. DGE1746891, and thanks David K. Sing and Leonardo A. dos Santos for interesting discussions surrounding this work. JMM acknowledges support from the Horizon Europe Guarantee Fund, grant EP/Z00330X/1. B.-O. D. acknowledges support from the Swiss State Secretariat for Education, Research and Innovation (SERI) under contract number MB22.00046. PCA acknowledges support from the Carlsberg Foundation, grant CF22-1254. EMV acknowledges financial support from the Swiss National Science Foundation (SNSF) Mobility Fellowship under grant no. P500PT\_225456/1. CEF acknowledges support from the European Research Council (ERC) under the European Union's Horizon 2020 research and innovation program under grant agreement no. 805445. NPG gratefully acknowledges support from Science Foundation Ireland and the Royal Society through a University Research Fellowship (URF\textbackslash R\textbackslash 201032). HJH acknowledges support from eSSENCE (grant number eSSENCE@LU 9:3), The Swedish National Research Council (project number 2023-05307), The Crafoord foundation and the Royal Physiographic Society of Lund, through The Fund of the Walter Gyllenberg Foundation. BP acknowledges support from the Walter Gyllenberg Foundation. ADR acknowledges support from the Carlsberg Foundation, grant CF22-1548.
\end{acknowledgments}

%

\facilities{JWST(MIRI)}


\software{astropy \citep{2013A&A...558A..33A,2018AJ....156..123A, AstropyCollaboration_2022}, batman \citep{batman}, corner \citep{corner}, dynesty \citep{dynesty}, george \citep{hodlr}, juliet \citep{juliet}, jupyter \citep{Kluyver2016jupyter}, matplotlib \citep{Hunter:2007}, multinest \citep{Feroz_2008, Feroz_2009, Feroz_2019}, NumPy \citep{harris2020array}, SciPy \citep{2020SciPy-NMeth}, seaborn \citep{seaborn}, transitspectroscopy \citep{transitspectroscopy}}



\pagebreak
\bibliography{bib}{}

\begin{thebibliography}{}
\expandafter\ifx\csname natexlab\endcsname\relax\def\natexlab#1{#1}\fi
\providecommand{\url}[1]{\href{#1}{#1}}
\providecommand{\dodoi}[1]{doi:~\href{http://doi.org/#1}{\nolinkurl{#1}}}
\providecommand{\doeprint}[1]{\href{http://ascl.net/#1}{\nolinkurl{http://ascl.net/#1}}}
\providecommand{\doarXiv}[1]{\href{https://arxiv.org/abs/#1}{\nolinkurl{https://arxiv.org/abs/#1}}}

\bibitem[{{Allard} {et~al.}(2011){Allard}, {Homeier}, \& {Freytag}}]{Allard_2011}
{Allard}, F., {Homeier}, D., \& {Freytag}, B. 2011, in Astronomical Society of the Pacific Conference Series, Vol. 448, 16th Cambridge Workshop on Cool Stars, Stellar Systems, and the Sun, ed. C.~{Johns-Krull}, M.~K. {Browning}, \& A.~A. {West}, 91, \dodoi{10.48550/arXiv.1011.5405}

\bibitem[{{Allard} {et~al.}(2012){Allard}, {Homeier}, \& {Freytag}}]{Allard_2012}
{Allard}, F., {Homeier}, D., \& {Freytag}, B. 2012, Philosophical Transactions of the Royal Society of London Series A, 370, 2765, \dodoi{10.1098/rsta.2011.0269}

\bibitem[{{Ambikasaran} {et~al.}(2015){Ambikasaran}, {Foreman-Mackey}, {Greengard}, {Hogg}, \& {O'Neil}}]{hodlr}
{Ambikasaran}, S., {Foreman-Mackey}, D., {Greengard}, L., {Hogg}, D.~W., \& {O'Neil}, M. 2015, IEEE Transactions on Pattern Analysis and Machine Intelligence, 38, 252, \dodoi{10.1109/TPAMI.2015.2448083}

\bibitem[{{Astropy Collaboration} {et~al.}(2013){Astropy Collaboration}, {Robitaille}, {Tollerud}, {Greenfield}, {Droettboom}, {Bray}, {Aldcroft}, {Davis}, {Ginsburg}, {Price-Whelan}, {Kerzendorf}, {Conley}, {Crighton}, {Barbary}, {Muna}, {Ferguson}, {Grollier}, {Parikh}, {Nair}, {Unther}, {Deil}, {Woillez}, {Conseil}, {Kramer}, {Turner}, {Singer}, {Fox}, {Weaver}, {Zabalza}, {Edwards}, {Azalee Bostroem}, {Burke}, {Casey}, {Crawford}, {Dencheva}, {Ely}, {Jenness}, {Labrie}, {Lim}, {Pierfederici}, {Pontzen}, {Ptak}, {Refsdal}, {Servillat}, \& {Streicher}}]{2013A&A...558A..33A}
{Astropy Collaboration}, {Robitaille}, T.~P., {Tollerud}, E.~J., {et~al.} 2013, \aap, 558, A33, \dodoi{10.1051/0004-6361/201322068}

\bibitem[{{Astropy Collaboration} {et~al.}(2018){Astropy Collaboration}, {Price-Whelan}, {Sip{\H{o}}cz}, {G{\"u}nther}, {Lim}, {Crawford}, {Conseil}, {Shupe}, {Craig}, {Dencheva}, {Ginsburg}, {VanderPlas}, {Bradley}, {P{\'e}rez-Su{\'a}rez}, {de Val-Borro}, {Aldcroft}, {Cruz}, {Robitaille}, {Tollerud}, {Ardelean}, {Babej}, {Bach}, {Bachetti}, {Bakanov}, {Bamford}, {Barentsen}, {Barmby}, {Baumbach}, {Berry}, {Biscani}, {Boquien}, {Bostroem}, {Bouma}, {Brammer}, {Bray}, {Breytenbach}, {Buddelmeijer}, {Burke}, {Calderone}, {Cano Rodr{\'\i}guez}, {Cara}, {Cardoso}, {Cheedella}, {Copin}, {Corrales}, {Crichton}, {D'Avella}, {Deil}, {Depagne}, {Dietrich}, {Donath}, {Droettboom}, {Earl}, {Erben}, {Fabbro}, {Ferreira}, {Finethy}, {Fox}, {Garrison}, {Gibbons}, {Goldstein}, {Gommers}, {Greco}, {Greenfield}, {Groener}, {Grollier}, {Hagen}, {Hirst}, {Homeier}, {Horton}, {Hosseinzadeh}, {Hu}, {Hunkeler}, {Ivezi{\'c}}, {Jain}, {Jenness}, {Kanarek}, {Kendrew}, {Kern}, {Kerzendorf}, {Khvalko}, {King}, {Kirkby}, {Kulkarni},
  {Kumar}, {Lee}, {Lenz}, {Littlefair}, {Ma}, {Macleod}, {Mastropietro}, {McCully}, {Montagnac}, {Morris}, {Mueller}, {Mumford}, {Muna}, {Murphy}, {Nelson}, {Nguyen}, {Ninan}, {N{\"o}the}, {Ogaz}, {Oh}, {Parejko}, {Parley}, {Pascual}, {Patil}, {Patil}, {Plunkett}, {Prochaska}, {Rastogi}, {Reddy Janga}, {Sabater}, {Sakurikar}, {Seifert}, {Sherbert}, {Sherwood-Taylor}, {Shih}, {Sick}, {Silbiger}, {Singanamalla}, {Singer}, {Sladen}, {Sooley}, {Sornarajah}, {Streicher}, {Teuben}, {Thomas}, {Tremblay}, {Turner}, {Terr{\'o}n}, {van Kerkwijk}, {de la Vega}, {Watkins}, {Weaver}, {Whitmore}, {Woillez}, {Zabalza}, \& {Astropy Contributors}}]{2018AJ....156..123A}
{Astropy Collaboration}, {Price-Whelan}, A.~M., {Sip{\H{o}}cz}, B.~M., {et~al.} 2018, \aj, 156, 123, \dodoi{10.3847/1538-3881/aabc4f}

\bibitem[{{Astropy Collaboration} {et~al.}(2022){Astropy Collaboration}, {Price-Whelan}, {Lim}, {Earl}, {Starkman}, {Bradley}, {Shupe}, {Patil}, {Corrales}, {Brasseur}, {N{\"o}the}, {Donath}, {Tollerud}, {Morris}, {Ginsburg}, {Vaher}, {Weaver}, {Tocknell}, {Jamieson}, {van Kerkwijk}, {Robitaille}, {Merry}, {Bachetti}, {G{\"u}nther}, {Aldcroft}, {Alvarado-Montes}, {Archibald}, {B{\'o}di}, {Bapat}, {Barentsen}, {Baz{\'a}n}, {Biswas}, {Boquien}, {Burke}, {Cara}, {Cara}, {Conroy}, {Conseil}, {Craig}, {Cross}, {Cruz}, {D'Eugenio}, {Dencheva}, {Devillepoix}, {Dietrich}, {Eigenbrot}, {Erben}, {Ferreira}, {Foreman-Mackey}, {Fox}, {Freij}, {Garg}, {Geda}, {Glattly}, {Gondhalekar}, {Gordon}, {Grant}, {Greenfield}, {Groener}, {Guest}, {Gurovich}, {Handberg}, {Hart}, {Hatfield-Dodds}, {Homeier}, {Hosseinzadeh}, {Jenness}, {Jones}, {Joseph}, {Kalmbach}, {Karamehmetoglu}, {Ka{\l}uszy{\'n}ski}, {Kelley}, {Kern}, {Kerzendorf}, {Koch}, {Kulumani}, {Lee}, {Ly}, {Ma}, {MacBride}, {Maljaars}, {Muna}, {Murphy}, {Norman},
  {O'Steen}, {Oman}, {Pacifici}, {Pascual}, {Pascual-Granado}, {Patil}, {Perren}, {Pickering}, {Rastogi}, {Roulston}, {Ryan}, {Rykoff}, {Sabater}, {Sakurikar}, {Salgado}, {Sanghi}, {Saunders}, {Savchenko}, {Schwardt}, {Seifert-Eckert}, {Shih}, {Jain}, {Shukla}, {Sick}, {Simpson}, {Singanamalla}, {Singer}, {Singhal}, {Sinha}, {Sip{\H{o}}cz}, {Spitler}, {Stansby}, {Streicher}, {{\v{S}}umak}, {Swinbank}, {Taranu}, {Tewary}, {Tremblay}, {de Val-Borro}, {Van Kooten}, {Vasovi{\'c}}, {Verma}, {de Miranda Cardoso}, {Williams}, {Wilson}, {Winkel}, {Wood-Vasey}, {Xue}, {Yoachim}, {Zhang}, {Zonca}, \& {Astropy Project Contributors}}]{AstropyCollaboration_2022}
{Astropy Collaboration}, {Price-Whelan}, A.~M., {Lim}, P.~L., {et~al.} 2022, \apj, 935, 167, \dodoi{10.3847/1538-4357/ac7c74}

\bibitem[{{August} {et~al.}(2025){August}, {Buchhave}, {Diamond-Lowe}, {Mendon{\c{c}}a}, {Gressier}, {Rathcke}, {Allen}, {Fortune}, {Jones}, {Meier Vald{\'e}s}, {Demory}, {Espinoza}, {Fisher}, {Gibson}, {Heng}, {Hoeijmakers}, {Hooton}, {Kitzmann}, {Prinoth}, {Eastman}, \& {Barnes}}]{August_2025}
{August}, P.~C., {Buchhave}, L.~A., {Diamond-Lowe}, H., {et~al.} 2025, \aap, 695, A171, \dodoi{10.1051/0004-6361/202452611}

\bibitem[{Batalha {et~al.}(2017)Batalha, Mandell, Pontoppidan, Stevenson, Lewis, Kalirai, Earl, Greene, Albert, \& Nielsen}]{pandexo}
Batalha, N.~E., Mandell, A., Pontoppidan, K., {et~al.} 2017, Publications of the Astronomical Society of the Pacific, 129, 064501, \dodoi{10.1088/1538-3873/aa65b0}

\bibitem[{{Bello-Arufe} {et~al.}(2025){Bello-Arufe}, {Damiano}, {Bennett}, {Hu}, {Welbanks}, {MacDonald}, {Seligman}, {Sing}, {Tokadjian}, {Oza}, \& {Yang}}]{Bello-Arufe_2025}
{Bello-Arufe}, A., {Damiano}, M., {Bennett}, K.~A., {et~al.} 2025, \apjl, 980, L26, \dodoi{10.3847/2041-8213/adaf22}

\bibitem[{{Bolmont} {et~al.}(2017){Bolmont}, {Selsis}, {Owen}, {Ribas}, {Raymond}, {Leconte}, \& {Gillon}}]{Bolmont_2017}
{Bolmont}, E., {Selsis}, F., {Owen}, J.~E., {et~al.} 2017, \mnras, 464, 3728, \dodoi{10.1093/mnras/stw2578}

\bibitem[{{Bonfanti} {et~al.}(2024){Bonfanti}, {Brady}, {Wilson}, {Venturini}, {Egger}, {Brandeker}, {Sousa}, {Lendl}, {Simon}, {Queloz}, {Olofsson}, {Adibekyan}, {Alibert}, {Fossati}, {Hooton}, {Kubyshkina}, {Luque}, {Murgas}, {Mustill}, {Santos}, {Van Grootel}, {Alonso}, {Asquier}, {Bandy}, {B{\'a}rczy}, {Barrado Navascues}, {Barros}, {Baumjohann}, {Bean}, {Beck}, {Beck}, {Benz}, {Bergomi}, {Billot}, {Borsato}, {Broeg}, {Collier Cameron}, {Csizmadia}, {Cubillos}, {Davies}, {Deleuil}, {Deline}, {Delrez}, {Demangeon}, {Demory}, {Ehrenreich}, {Erikson}, {Fortier}, {Fridlund}, {Gandolfi}, {Gillon}, {G{\"u}del}, {G{\"u}nther}, {Heitzmann}, {Helling}, {Hoyer}, {Isaak}, {Kasper}, {Kiss}, {Lam}, {Laskar}, {Lecavelier des Etangs}, {Magrin}, {Maxted}, {Mordasini}, {Nascimbeni}, {Ottensamer}, {Pagano}, {Pall{\'e}}, {Peter}, {Piotto}, {Pollacco}, {Ragazzoni}, {Rando}, {Rauer}, {Ribas}, {Scandariato}, {S{\'e}gransan}, {Seifahrt}, {Smith}, {Stalport}, {Stef{\'a}nsson}, {Steinberger}, {St{\"u}rmer}, {Szab{\'o}}, {Thomas},
  {Udry}, {Villaver}, {Walton}, {Westerdorff}, \& {Zingales}}]{Bonfanti_2024}
{Bonfanti}, A., {Brady}, M., {Wilson}, T.~G., {et~al.} 2024, \aap, 682, A66, \dodoi{10.1051/0004-6361/202348180}

\bibitem[{{Bouwman} {et~al.}(2023){Bouwman}, {Kendrew}, {Greene}, {Bell}, {Lagage}, {Schreiber}, {Dicken}, {Sloan}, {Espinoza}, {Scheithauer}, {Coulais}, {Fox}, {Gastaud}, {Glauser}, {Jones}, {Labiano}, {Lahuis}, {Morrison}, {Murray}, {Mueller}, {Nayak}, {Wright}, {Glasse}, \& {Rieke}}]{Bouwman_2023}
{Bouwman}, J., {Kendrew}, S., {Greene}, T.~P., {et~al.} 2023, \pasp, 135, 038002, \dodoi{10.1088/1538-3873/acbc49}

\bibitem[{Bradley {et~al.}(2024)Bradley, Sip{\H o}cz, Robitaille, Tollerud, Vin{\'{\i}}cius, Deil, Barbary, Wilson, Busko, Donath, G{\"u}nther, Cara, Lim, Me{\ss}linger, Burnett, Conseil, Droettboom, Bostroem, Bray, Bratholm, Jamieson, Ginsburg, Barentsen, Craig, Pascual, Rathi, Perrin, Morris, \& Perren}]{photutil}
Bradley, L., Sip{\H o}cz, B., Robitaille, T., {et~al.} 2024, astropy/photutils: 1.12.0, 1.12.0,  Zenodo, \dodoi{10.5281/zenodo.10967176}

\bibitem[{{Brunetto} {et~al.}(2015){Brunetto}, {Loeffler}, {Nesvorn{\'y}}, {Sasaki}, \& {Strazzulla}}]{Brunetto_2015}
{Brunetto}, R., {Loeffler}, M.~J., {Nesvorn{\'y}}, D., {Sasaki}, S., \& {Strazzulla}, G. 2015, in Asteroids IV, ed. P.~{Michel}, F.~E. {DeMeo}, \& W.~F. {Bottke}, 597--616, \dodoi{10.2458/azu_uapress_9780816532131-ch031}

\bibitem[{{Bushouse} {et~al.}(2023){Bushouse}, {Eisenhamer}, {Dencheva}, {Davies}, {Greenfield}, {Morrison}, {Hodge}, {Simon}, {Grumm}, {Droettboom}, {Slavich}, {Sosey}, {Pauly}, {Miller}, {Jedrzejewski}, {Hack}, {Davis}, {Crawford}, {Law}, {Gordon}, {Regan}, {Cara}, {MacDonald}, {Bradley}, {Shanahan}, {Jamieson}, {Teodoro}, \& {Williams}}]{Bushouse_2023}
{Bushouse}, H., {Eisenhamer}, J., {Dencheva}, N., {et~al.} 2023, {JWST Calibration Pipeline}, 1.9.4,  Zenodo, \dodoi{10.5281/zenodo.7577320}

\bibitem[{Catling \& Kasting(2007)}]{catling_2007}
Catling, D., \& Kasting, J. 2007, Planetary atmospheres and life (United Kingdom: Cambridge University Press), 91--116

\bibitem[{{Cloutier} {et~al.}(2020){Cloutier}, {Eastman}, {Rodriguez}, {Astudillo-Defru}, {Bonfils}, {Mortier}, {Watson}, {Stalport}, {Pinamonti}, {Lienhard}, {Harutyunyan}, {Damasso}, {Latham}, {Collins}, {Massey}, {Irwin}, {Winters}, {Charbonneau}, {Ziegler}, {Matthews}, {Crossfield}, {Kreidberg}, {Quinn}, {Ricker}, {Vanderspek}, {Seager}, {Winn}, {Jenkins}, {Vezie}, {Udry}, {Twicken}, {Tenenbaum}, {Sozzetti}, {S{\'e}gransan}, {Schlieder}, {Sasselov}, {Santos}, {Rice}, {Rackham}, {Poretti}, {Piotto}, {Phillips}, {Pepe}, {Molinari}, {Mignon}, {Micela}, {Melo}, {de Medeiros}, {Mayor}, {Matson}, {Martinez Fiorenzano}, {Mann}, {Magazz{\'u}}, {Lovis}, {L{\'o}pez-Morales}, {Lopez}, {Lissauer}, {L{\'e}pine}, {Law}, {Kielkopf}, {Johnson}, {Jensen}, {Howell}, {Gonzales}, {Ghedina}, {Forveille}, {Figueira}, {Dumusque}, {Dressing}, {Doyon}, {D{\'\i}az}, {Fabrizio}, {Delfosse}, {Cosentino}, {Conti}, {Collins}, {Cameron}, {Ciardi}, {Caldwell}, {Burke}, {Buchhave}, {Brice{\~n}o}, {Boyd}, {Bouchy}, {Beichman}, {Artigau},
  \& {Almenara}}]{Cloutier_2020}
{Cloutier}, R., {Eastman}, J.~D., {Rodriguez}, J.~E., {et~al.} 2020, \aj, 160, 3, \dodoi{10.3847/1538-3881/ab91c2}

\bibitem[{{Crossfield} {et~al.}(2022){Crossfield}, {Malik}, {Hill}, {Kane}, {Foley}, {Polanski}, {Coria}, {Brande}, {Zhang}, {Wienke}, {Kreidberg}, {Cowan}, {Dragomir}, {Gorjian}, {Mikal-Evans}, {Benneke}, {Christiansen}, {Deming}, \& {Morales}}]{Crossfield_2022}
{Crossfield}, I. J.~M., {Malik}, M., {Hill}, M.~L., {et~al.} 2022, \apjl, 937, L17, \dodoi{10.3847/2041-8213/ac886b}

\bibitem[{Deal \& Espinoza(2024)}]{spelunker}
Deal, D., \& Espinoza, N. 2024, Journal of Open Source Software, 9, 6202, \dodoi{10.21105/joss.06202}

\bibitem[{{Demory} {et~al.}(2023){Demory}, {Sulis}, {Meier Vald{\'e}s}, {Delrez}, {Brandeker}, {Billot}, {Fortier}, {Hoyer}, {Sousa}, {Heng}, {Lendl}, {Krenn}, {Morris}, {Patel}, {Alibert}, {Alonso}, {Anglada}, {B{\'a}rczy}, {Barrado}, {Barros}, {Baumjohann}, {Beck}, {Beck}, {Benz}, {Bonfils}, {Broeg}, {Buder}, {Cabrera}, {Charnoz}, {Collier Cameron}, {Cottard}, {Csizmadia}, {Davies}, {Deleuil}, {Demangeon}, {Ehrenreich}, {Erikson}, {Fossati}, {Fridlund}, {Gandolfi}, {Gillon}, {G{\"u}del}, {Isaak}, {Kiss}, {Laskar}, {Lecavelier des Etangs}, {Lovis}, {Luntzer}, {Magrin}, {Marafatto}, {Maxted}, {Nascimbeni}, {Olofsson}, {Ottensamer}, {Pagano}, {Pall{\'e}}, {Peter}, {Piotto}, {Pollacco}, {Queloz}, {Ragazzoni}, {Rando}, {Ratti}, {Rauer}, {Ribas}, {Santos}, {Scandariato}, {S{\'e}gransan}, {Simon}, {Smith}, {Steller}, {Szab{\'o}}, {Thomas}, {Udry}, {Van Grootel}, \& {Walton}}]{Demory_2023}
{Demory}, B.~O., {Sulis}, S., {Meier Vald{\'e}s}, E., {et~al.} 2023, \aap, 669, A64, \dodoi{10.1051/0004-6361/202244894}

\bibitem[{{Domingue} {et~al.}(2014){Domingue}, {Chapman}, {Killen}, {Zurbuchen}, {Gilbert}, {Sarantos}, {Benna}, {Slavin}, {Schriver}, {Tr{\'a}vn{\'\i}{\v{c}}ek}, {Orlando}, {Sprague}, {Blewett}, {Gillis-Davis}, {Feldman}, {Lawrence}, {Ho}, {Ebel}, {Nittler}, {Vilas}, {Pieters}, {Solomon}, {Johnson}, {Winslow}, {Helbert}, {Peplowski}, {Weider}, {Mouawad}, {Izenberg}, \& {McClintock}}]{Domingue_2014}
{Domingue}, D.~L., {Chapman}, C.~R., {Killen}, R.~M., {et~al.} 2014, \ssr, 181, 121, \dodoi{10.1007/s11214-014-0039-5}

\bibitem[{{Dorn} {et~al.}(2018){Dorn}, {Noack}, \& {Rozel}}]{Dorn_2018}
{Dorn}, C., {Noack}, L., \& {Rozel}, A.~B. 2018, \aap, 614, A18, \dodoi{10.1051/0004-6361/201731513}

\bibitem[{{Ducrot} {et~al.}(2025){Ducrot}, {Lagage}, {Min}, {Gillon}, {Bell}, {Tremblin}, {Greene}, {Dyrek}, {Bouwman}, {Waters}, {G{\"u}del}, {Henning}, {Vandenbussche}, {Absil}, {Barrado}, {Boccaletti}, {Coulais}, {Decin}, {Edwards}, {Gastaud}, {Glasse}, {Kendrew}, {Olofsson}, {Patapis}, {Pye}, {Rouan}, {Whiteford}, {Argyriou}, {Cossou}, {Glauser}, {Krause}, {Lahuis}, {Royer}, {Scheithauer}, {Colina}, {van Dishoeck}, {Ostlin}, {Ray}, \& {Wright}}]{Ducrot_2025}
{Ducrot}, E., {Lagage}, P.-O., {Min}, M., {et~al.} 2025, Nature Astronomy, 9, 358, \dodoi{10.1038/s41550-024-02428-z}

\bibitem[{{Eastman} {et~al.}(2013){Eastman}, {Gaudi}, \& {Agol}}]{Eastman_2013}
{Eastman}, J., {Gaudi}, B.~S., \& {Agol}, E. 2013, \pasp, 125, 83, \dodoi{10.1086/669497}

\bibitem[{{Engle}(2024)}]{Engle_2024}
{Engle}, S.~G. 2024, \apj, 960, 62, \dodoi{10.3847/1538-4357/ad0840}

\bibitem[{Espinoza(2022)}]{transitspectroscopy}
Espinoza, N. 2022, TransitSpectroscopy, 0.3.11,  Zenodo, \dodoi{10.5281/zenodo.6960924}

\bibitem[{{Espinoza} {et~al.}(2019){Espinoza}, {Kossakowski}, \& {Brahm}}]{juliet}
{Espinoza}, N., {Kossakowski}, D., \& {Brahm}, R. 2019, \mnras, 490, 2262, \dodoi{10.1093/mnras/stz2688}

\bibitem[{Feroz \& Hobson(2008)}]{Feroz_2008}
Feroz, F., \& Hobson, M.~P. 2008, Monthly Notices of the Royal Astronomical Society, 384, 449, \dodoi{10.1111/j.1365-2966.2007.12353.x}

\bibitem[{Feroz {et~al.}(2009)Feroz, Hobson, \& Bridges}]{Feroz_2009}
Feroz, F., Hobson, M.~P., \& Bridges, M. 2009, Monthly Notices of the Royal Astronomical Society, 398, 1601, \dodoi{10.1111/j.1365-2966.2009.14548.x}

\bibitem[{Feroz {et~al.}(2019)Feroz, Hobson, Cameron, \& Pettitt}]{Feroz_2019}
Feroz, F., Hobson, M.~P., Cameron, E., \& Pettitt, A.~N. 2019, The Open Journal of Astrophysics, 2, 10, \dodoi{10.21105/astro.1306.2144}

\bibitem[{Foreman-Mackey(2016)}]{corner}
Foreman-Mackey, D. 2016, The Journal of Open Source Software, 1, 24, \dodoi{10.21105/joss.00024}

\bibitem[{Foreman-Mackey {et~al.}(2017)Foreman-Mackey, Agol, Ambikasaran, \& Angus}]{celerite}
Foreman-Mackey, D., Agol, E., Ambikasaran, S., \& Angus, R. 2017, The Astronomical Journal, 154, 220, \dodoi{10.3847/1538-3881/aa9332}

\bibitem[{{France} {et~al.}(2016){France}, {Loyd}, {Youngblood}, {Brown}, {Schneider}, {Hawley}, {Froning}, {Linsky}, {Roberge}, {Buccino}, {Davenport}, {Fontenla}, {Kaltenegger}, {Kowalski}, {Mauas}, {Miguel}, {Redfield}, {Rugheimer}, {Tian}, {Vieytes}, {Walkowicz}, \& {Weisenburger}}]{France_2016}
{France}, K., {Loyd}, R.~O.~P., {Youngblood}, A., {et~al.} 2016, \apj, 820, 89, \dodoi{10.3847/0004-637X/820/2/89}

\bibitem[{{Gaillard} \& {Scaillet}(2014)}]{Gaillard_2014}
{Gaillard}, F., \& {Scaillet}, B. 2014, Earth and Planetary Science Letters, 403, 307, \dodoi{10.1016/j.epsl.2014.07.009}

\bibitem[{{Gibson} {et~al.}(2012){Gibson}, {Aigrain}, {Roberts}, {Evans}, {Osborne}, \& {Pont}}]{Gibson_2012}
{Gibson}, N.~P., {Aigrain}, S., {Roberts}, S., {et~al.} 2012, \mnras, 419, 2683, \dodoi{10.1111/j.1365-2966.2011.19915.x}

\bibitem[{{Gillmann} {et~al.}(2024){Gillmann}, {Arney}, {Avice}, {Dyar}, {Golabek}, {G{\"u}lcher}, {Johnson}, {Lefevre}, \& {Widemann}}]{Gillmann_2024}
{Gillmann}, C., {Arney}, G.~N., {Avice}, G., {et~al.} 2024, arXiv e-prints, arXiv:2404.07669, \dodoi{10.48550/arXiv.2404.07669}

\bibitem[{{Gillon} {et~al.}(2017){Gillon}, {Triaud}, {Demory}, {Jehin}, {Agol}, {Deck}, {Lederer}, {de Wit}, {Burdanov}, {Ingalls}, {Bolmont}, {Leconte}, {Raymond}, {Selsis}, {Turbet}, {Barkaoui}, {Burgasser}, {Burleigh}, {Carey}, {Chaushev}, {Copperwheat}, {Delrez}, {Fernandes}, {Holdsworth}, {Kotze}, {Van Grootel}, {Almleaky}, {Benkhaldoun}, {Magain}, \& {Queloz}}]{Gillon_2017}
{Gillon}, M., {Triaud}, A. H.~M.~J., {Demory}, B.-O., {et~al.} 2017, \nat, 542, 456, \dodoi{10.1038/nature21360}

\bibitem[{{Gordon} {et~al.}(2025){Gordon}, {Sloan}, {Garcia Marin}, {Libralato}, {Rieke}, {Aguilar}, {Bohlin}, {Cracraft}, {Decleir}, {Gaspar}, {Kendrew}, {Law}, {Noriega-Crespo}, \& {Regan}}]{Gordon_2025}
{Gordon}, K.~D., {Sloan}, G.~C., {Garcia Marin}, M., {et~al.} 2025, \aj, 169, 6, \dodoi{10.3847/1538-3881/ad8cd4}

\bibitem[{{Greene} {et~al.}(2023){Greene}, {Bell}, {Ducrot}, {Dyrek}, {Lagage}, \& {Fortney}}]{greene_2023}
{Greene}, T.~P., {Bell}, T.~J., {Ducrot}, E., {et~al.} 2023, \nat, 618, 39, \dodoi{10.1038/s41586-023-05951-7}

\bibitem[{{Gressier} {et~al.}(2024){Gressier}, {Espinoza}, {Allen}, {Sing}, {Banerjee}, {Barstow}, {Valenti}, {Lewis}, {Birkmann}, {Challener}, {Manjavacas}, {Alves de Oliveira}, {Crouzet}, \& {Beck}}]{Gressier_2024}
{Gressier}, A., {Espinoza}, N., {Allen}, N.~H., {et~al.} 2024, \apjl, 975, L10, \dodoi{10.3847/2041-8213/ad73d1}

\bibitem[{{Hammond} {et~al.}(2025){Hammond}, {Guimond}, {Lichtenberg}, {Nicholls}, {Fisher}, {Luque}, {Meier}, {Taylor}, {Changeat}, {Dang}, {Hay}, {Herbort}, \& {Teske}}]{Hammond_2025}
{Hammond}, M., {Guimond}, C.~M., {Lichtenberg}, T., {et~al.} 2025, \apjl, 978, L40, \dodoi{10.3847/2041-8213/ada0bc}

\bibitem[{{Hansen}(2008)}]{Hansen_2008}
{Hansen}, B. M.~S. 2008, \apjs, 179, 484, \dodoi{10.1086/591964}

\bibitem[{Harris {et~al.}(2020)Harris, Millman, van~der Walt, Gommers, Virtanen, Cournapeau, Wieser, Taylor, Berg, Smith, Kern, Picus, Hoyer, van Kerkwijk, Brett, Haldane, del R{\'{i}}o, Wiebe, Peterson, G{\'{e}}rard-Marchant, Sheppard, Reddy, Weckesser, Abbasi, Gohlke, \& Oliphant}]{harris2020array}
Harris, C.~R., Millman, K.~J., van~der Walt, S.~J., {et~al.} 2020, Nature, 585, 357, \dodoi{10.1038/s41586-020-2649-2}

\bibitem[{{Herbort} {et~al.}(2020){Herbort}, {Woitke}, {Helling}, \& {Zerkle}}]{Herbort_2020}
{Herbort}, O., {Woitke}, P., {Helling}, C., \& {Zerkle}, A. 2020, \aap, 636, A71, \dodoi{10.1051/0004-6361/201936614}

\bibitem[{Hinne {et~al.}(2020)Hinne, Gronau, van~den Bergh, \& Wagenmakers}]{Hinne_2020}
Hinne, M., Gronau, Q.~F., van~den Bergh, D., \& Wagenmakers, E.-J. 2020, Advances in Methods and Practices in Psychological Science, 3, 200–215, \dodoi{10.1177/2515245919898657}

\bibitem[{{Hu} {et~al.}(2023){Hu}, {Gaillard}, \& {Kite}}]{Hu_2023}
{Hu}, R., {Gaillard}, F., \& {Kite}, E.~S. 2023, \apjl, 948, L20, \dodoi{10.3847/2041-8213/acd0b4}

\bibitem[{{Hu} {et~al.}(2024){Hu}, {Bello-Arufe}, {Zhang}, {Paragas}, {Zilinskas}, {van Buchem}, {Bess}, {Patel}, {Ito}, {Damiano}, {Scheucher}, {Oza}, {Knutson}, {Miguel}, {Dragomir}, {Brandeker}, \& {Demory}}]{Hu_2024}
{Hu}, R., {Bello-Arufe}, A., {Zhang}, M., {et~al.} 2024, \nat, 630, 609, \dodoi{10.1038/s41586-024-07432-x}

\bibitem[{Hunter(2007)}]{Hunter:2007}
Hunter, J.~D. 2007, Computing in Science \& Engineering, 9, 90, \dodoi{10.1109/MCSE.2007.55}

\bibitem[{{Ih} {et~al.}(2023){Ih}, {Kempton}, {Whittaker}, \& {Lessard}}]{Ih_2023}
{Ih}, J., {Kempton}, E. M.~R., {Whittaker}, E.~A., \& {Lessard}, M. 2023, \apjl, 952, L4, \dodoi{10.3847/2041-8213/ace03b}

\bibitem[{{Johnstone} {et~al.}(2021){Johnstone}, {Bartel}, \& {G{\"u}del}}]{Johnstone_2021}
{Johnstone}, C.~P., {Bartel}, M., \& {G{\"u}del}, M. 2021, \aap, 649, A96, \dodoi{10.1051/0004-6361/202038407}

\bibitem[{{Kirk} {et~al.}(2024){Kirk}, {Stevenson}, {Fu}, {Lustig-Yaeger}, {Moran}, {Peacock}, {Alam}, {Batalha}, {Bennett}, {Gonzalez-Quiles}, {L{\'o}pez-Morales}, {Lothringer}, {MacDonald}, {May}, {Mayorga}, {Rustamkulov}, {Sing}, {Sotzen}, {Valenti}, \& {Wakeford}}]{kirk_2024}
{Kirk}, J., {Stevenson}, K.~B., {Fu}, G., {et~al.} 2024, \aj, 167, 90, \dodoi{10.3847/1538-3881/ad19df}

\bibitem[{Kluyver {et~al.}(2016)Kluyver, Ragan-Kelley, P{\'e}rez, Granger, Bussonnier, Frederic, Kelley, Hamrick, Grout, Corlay, Ivanov, Avila, Abdalla, \& Willing}]{Kluyver2016jupyter}
Kluyver, T., Ragan-Kelley, B., P{\'e}rez, F., {et~al.} 2016, in Positioning and Power in Academic Publishing: Players, Agents and Agendas, ed. F.~Loizides \& B.~Schmidt, IOS Press, 87 -- 90

\bibitem[{{Koll}(2022)}]{Koll_2022}
{Koll}, D. D.~B. 2022, \apj, 924, 134, \dodoi{10.3847/1538-4357/ac3b48}

\bibitem[{{Kreidberg}(2015)}]{batman}
{Kreidberg}, L. 2015, \pasp, 127, 1161, \dodoi{10.1086/683602}

\bibitem[{{Kreidberg} {et~al.}(2019){Kreidberg}, {Koll}, {Morley}, {Hu}, {Schaefer}, {Deming}, {Stevenson}, {Dittmann}, {Vanderburg}, {Berardo}, {Guo}, {Stassun}, {Crossfield}, {Charbonneau}, {Latham}, {Loeb}, {Ricker}, {Seager}, \& {Vanderspek}}]{kreidberg_2019}
{Kreidberg}, L., {Koll}, D. D.~B., {Morley}, C., {et~al.} 2019, \nat, 573, 87, \dodoi{10.1038/s41586-019-1497-4}

\bibitem[{{Krissansen-Totton} {et~al.}(2024){Krissansen-Totton}, {Wogan}, {Thompson}, \& {Fortney}}]{Krissansen-Totton_2024}
{Krissansen-Totton}, J., {Wogan}, N., {Thompson}, M., \& {Fortney}, J.~J. 2024, Nature Communications, 15, 8374, \dodoi{10.1038/s41467-024-52642-6}

\bibitem[{{Laurent} {et~al.}(2020){Laurent}, {Bj{\"o}rnsen}, {Wotzlaw}, {Bretscher}, {Pimenta Silva}, {Moyen}, {Ulmer}, \& {Bachmann}}]{Laurent_2020}
{Laurent}, O., {Bj{\"o}rnsen}, J., {Wotzlaw}, J.-F., {et~al.} 2020, Nature Geoscience, 13, 163, \dodoi{10.1038/s41561-019-0520-6}

\bibitem[{{Libralato} {et~al.}(2024){Libralato}, {Argyriou}, {Dicken}, {Garc{\'\i}a Mar{\'\i}n}, {Guillard}, {Hines}, {Kavanagh}, {Kendrew}, {Law}, {Noriega-Crespo}, \& {{\'A}lvarez-M{\'a}rquez}}]{Libralato_2024}
{Libralato}, M., {Argyriou}, I., {Dicken}, D., {et~al.} 2024, \pasp, 136, 034502, \dodoi{10.1088/1538-3873/ad2551}

\bibitem[{{Lim} {et~al.}(2023){Lim}, {Benneke}, {Doyon}, {MacDonald}, {Piaulet}, {Artigau}, {Coulombe}, {Radica}, {L'Heureux}, {Albert}, {Rackham}, {de Wit}, {Salhi}, {Roy}, {Flagg}, {Fournier-Tondreau}, {Taylor}, {Cook}, {Lafreni{\`e}re}, {Cowan}, {Kaltenegger}, {Rowe}, {Espinoza}, {Dang}, \& {Darveau-Bernier}}]{lim_2023}
{Lim}, O., {Benneke}, B., {Doyon}, R., {et~al.} 2023, \apjl, 955, L22, \dodoi{10.3847/2041-8213/acf7c4}

\bibitem[{{Luger} {et~al.}(2019){Luger}, {Agol}, {Foreman-Mackey}, {Fleming}, {Lustig-Yaeger}, \& {Deitrick}}]{Luger_2019}
{Luger}, R., {Agol}, E., {Foreman-Mackey}, D., {et~al.} 2019, \aj, 157, 64, \dodoi{10.3847/1538-3881/aae8e5}

\bibitem[{{Luger} \& {Barnes}(2015)}]{Luger_2015}
{Luger}, R., \& {Barnes}, R. 2015, Astrobiology, 15, 119, \dodoi{10.1089/ast.2014.1231}

\bibitem[{{Luque} \& {Pall{\'e}}(2022)}]{Luque_2022}
{Luque}, R., \& {Pall{\'e}}, E. 2022, Science, 377, 1211, \dodoi{10.1126/science.abl7164}

\bibitem[{{Luque} {et~al.}(2024){Luque}, {Park Coy}, {Xue}, {Feinstein}, {Ahrer}, {Changeat}, {Zhang}, {Moran}, {Bean}, {Kite}, {Weiner Mansfield}, \& {Pall{\'e}}}]{Luque_2024}
{Luque}, R., {Park Coy}, B., {Xue}, Q., {et~al.} 2024, arXiv e-prints, arXiv:2412.03411, \dodoi{10.48550/arXiv.2412.03411}

\bibitem[{{Lustig-Yaeger} {et~al.}(2023){Lustig-Yaeger}, {Fu}, {May}, {Ceballos}, {Moran}, {Peacock}, {Stevenson}, {Kirk}, {L{\'o}pez-Morales}, {MacDonald}, {Mayorga}, {Sing}, {Sotzen}, {Valenti}, {Redai}, {Alam}, {Batalha}, {Bennett}, {Gonzalez-Quiles}, {Kruse}, {Lothringer}, {Rustamkulov}, \& {Wakeford}}]{lustig-yaeger_2023}
{Lustig-Yaeger}, J., {Fu}, G., {May}, E.~M., {et~al.} 2023, Nature Astronomy, 7, 1317, \dodoi{10.1038/s41550-023-02064-z}

\bibitem[{{Malik} {et~al.}(2019{\natexlab{a}}){Malik}, {Kempton}, {Koll}, {Mansfield}, {Bean}, \& {Kite}}]{Malik_2019b}
{Malik}, M., {Kempton}, E. M.~R., {Koll}, D. D.~B., {et~al.} 2019{\natexlab{a}}, The Astrophysical Journal, 886, 142, \dodoi{10.3847/1538-4357/ab4a05}

\bibitem[{{Malik} {et~al.}(2019{\natexlab{b}}){Malik}, {Kitzmann}, {Mendon{\c{c}}a}, {Grimm}, {Marleau}, {Linder}, {Tsai}, \& {Heng}}]{Malik_2019a}
{Malik}, M., {Kitzmann}, D., {Mendon{\c{c}}a}, J.~M., {et~al.} 2019{\natexlab{b}}, The Astronomical Journal, 157, 170, \dodoi{10.3847/1538-3881/ab1084}

\bibitem[{{Malik} {et~al.}(2017){Malik}, {Grosheintz}, {Mendon{\c{c}}a}, {Grimm}, {Lavie}, {Kitzmann}, {Tsai}, {Burrows}, {Kreidberg}, {Bedell}, {Bean}, {Stevenson}, \& {Heng}}]{Malik_2017}
{Malik}, M., {Grosheintz}, L., {Mendon{\c{c}}a}, J.~M., {et~al.} 2017, The Astronomical Journal, 153, 56, \dodoi{10.3847/1538-3881/153/2/56}

\bibitem[{{Mansfield} {et~al.}(2019){Mansfield}, {Kite}, {Hu}, {Koll}, {Malik}, {Bean}, \& {Kempton}}]{Mansfield_2019}
{Mansfield}, M., {Kite}, E.~S., {Hu}, R., {et~al.} 2019, \apj, 886, 141, \dodoi{10.3847/1538-4357/ab4c90}

\bibitem[{{May} {et~al.}(2023){May}, {MacDonald}, {Bennett}, {Moran}, {Wakeford}, {Peacock}, {Lustig-Yaeger}, {Highland}, {Stevenson}, {Sing}, {Mayorga}, {Batalha}, {Kirk}, {L{\'o}pez-Morales}, {Valenti}, {Alam}, {Alderson}, {Fu}, {Gonzalez-Quiles}, {Lothringer}, {Rustamkulov}, \& {Sotzen}}]{may_2023}
{May}, E.~M., {MacDonald}, R.~J., {Bennett}, K.~A., {et~al.} 2023, \apjl, 959, L9, \dodoi{10.3847/2041-8213/ad054f}

\bibitem[{Meier~Valdés {et~al.}(2025)Meier~Valdés, Demory, Diamond-Lowe, Mendonça, August, Fortune, Allen, Kitzmann, Gressier, Hooton, Jones, Buchhave, Espinoza, Fisher, Gibson, Heng, Hoeijmakers, Prinoth, Rathcke, \& Eastman}]{Meier_Valdes_2025}
Meier~Valdés, E., Demory, B.-O., Diamond-Lowe, H., {et~al.} 2025, Astronomy \& Astrophysics, \dodoi{10.1051/0004-6361/202453449}

\bibitem[{{Moore} \& {Cowan}(2020)}]{Moore_2020}
{Moore}, K., \& {Cowan}, N.~B. 2020, \mnras, 496, 3786, \dodoi{10.1093/mnras/staa1796}

\bibitem[{{Moran} {et~al.}(2023){Moran}, {Stevenson}, {Sing}, {MacDonald}, {Kirk}, {Lustig-Yaeger}, {Peacock}, {Mayorga}, {Bennett}, {L{\'o}pez-Morales}, {May}, {Rustamkulov}, {Valenti}, {Adams Redai}, {Alam}, {Batalha}, {Fu}, {Gonzalez-Quiles}, {Highland}, {Kruse}, {Lothringer}, {Ortiz Ceballos}, {Sotzen}, \& {Wakeford}}]{moran_2023}
{Moran}, S.~E., {Stevenson}, K.~B., {Sing}, D.~K., {et~al.} 2023, \apjl, 948, L11, \dodoi{10.3847/2041-8213/accb9c}

\bibitem[{{Morrison} {et~al.}(2023){Morrison}, {Dicken}, {Argyriou}, {Ressler}, {Gordon}, {Regan}, {Cracraft}, {Rieke}, {Engesser}, {Alberts}, {Alvarez-Marquez}, {Colbert}, {Fox}, {Gasman}, {Law}, {Garcia Marin}, {G{\'a}sp{\'a}r}, {Guillard}, {Kendrew}, {Labiano}, {Laine}, {Noriega-Crespo}, {Shivaei}, \& {Sloan}}]{Morrison_2023}
{Morrison}, J.~E., {Dicken}, D., {Argyriou}, I., {et~al.} 2023, \pasp, 135, 075004, \dodoi{10.1088/1538-3873/acdea6}

\bibitem[{{Nowak} {et~al.}(2020){Nowak}, {Luque}, {Parviainen}, {Pall{\'e}}, {Molaverdikhani}, {B{\'e}jar}, {Lillo-Box}, {Rodr{\'\i}guez-L{\'o}pez}, {Caballero}, {Zechmeister}, {Passegger}, {Cifuentes}, {Schweitzer}, {Narita}, {Cale}, {Espinoza}, {Murgas}, {Hidalgo}, {Zapatero Osorio}, {Pozuelos}, {Aceituno}, {Amado}, {Barkaoui}, {Barrado}, {Bauer}, {Benkhaldoun}, {Caldwell}, {Casasayas Barris}, {Chaturvedi}, {Chen}, {Collins}, {Collins}, {Cort{\'e}s-Contreras}, {Crossfield}, {de Le{\'o}n}, {D{\'\i}ez Alonso}, {Dreizler}, {El Mufti}, {Esparza-Borges}, {Essack}, {Fukui}, {Gaidos}, {Gillon}, {Gonzales}, {Guerra}, {Hatzes}, {Henning}, {Herrero}, {Hesse}, {Hirano}, {Howell}, {Jeffers}, {Jehin}, {Jenkins}, {Kaminski}, {Kemmer}, {Kielkopf}, {Kossakowski}, {Kotani}, {K{\"u}rster}, {Lafarga}, {Latham}, {Law}, {Lissauer}, {Lodieu}, {Madrigal-Aguado}, {Mann}, {Massey}, {Matson}, {Matthews}, {Monta{\~n}{\'e}s-Rodr{\'\i}guez}, {Montes}, {Morales}, {Mori}, {Nagel}, {Oshagh}, {Pedraz}, {Plavchan}, {Pollacco},
  {Quirrenbach}, {Reffert}, {Reiners}, {Ribas}, {Ricker}, {Rose}, {Schlecker}, {Schlieder}, {Seager}, {Stangret}, {Stock}, {Tamura}, {Tanner}, {Teske}, {Trifonov}, {Twicken}, {Vanderspek}, {Watanabe}, {Wittrock}, {Ziegler}, \& {Zohrabi}}]{Nowak_2020}
{Nowak}, G., {Luque}, R., {Parviainen}, H., {et~al.} 2020, \aap, 642, A173, \dodoi{10.1051/0004-6361/202037867}

\bibitem[{{Paragas} {et~al.}(2025){Paragas}, {Knutson}, {Hu}, {Ehlmann}, {Alemanno}, {Helbert}, {Maturilli}, {Zhang}, {Iyer}, \& {Rossman}}]{Paragas_2025}
{Paragas}, K., {Knutson}, H.~A., {Hu}, R., {et~al.} 2025, \apj, 981, 130, \dodoi{10.3847/1538-4357/ada9eb}

\bibitem[{{Pass} {et~al.}(2025){Pass}, {Charbonneau}, \& {Vanderburg}}]{Pass_2025}
{Pass}, E.~K., {Charbonneau}, D., \& {Vanderburg}, A. 2025, arXiv e-prints, arXiv:2504.01182, \dodoi{10.48550/arXiv.2504.01182}

\bibitem[{{Patel} {et~al.}(2024){Patel}, {Brandeker}, {Kitzmann}, {Petit dit de la Roche}, {Bello-Arufe}, {Heng}, {Meier Vald{\'e}s}, {Persson}, {Zhang}, {Demory}, {Bourrier}, {Deline}, {Ehrenreich}, {Fridlund}, {Hu}, {Lendl}, {Oza}, {Alibert}, \& {Hooton}}]{Patel_2024}
{Patel}, J.~A., {Brandeker}, A., {Kitzmann}, D., {et~al.} 2024, \aap, 690, A159, \dodoi{10.1051/0004-6361/202450748}

\bibitem[{{Perrin} {et~al.}(2012){Perrin}, {Soummer}, {Elliott}, {Lallo}, \& {Sivaramakrishnan}}]{Perrin_2012}
{Perrin}, M.~D., {Soummer}, R., {Elliott}, E.~M., {Lallo}, M.~D., \& {Sivaramakrishnan}, A. 2012, in Society of Photo-Optical Instrumentation Engineers (SPIE) Conference Series, Vol. 8442, Space Telescopes and Instrumentation 2012: Optical, Infrared, and Millimeter Wave, ed. M.~C. {Clampin}, G.~G. {Fazio}, H.~A. {MacEwen}, \& J.~{Oschmann}, Jacobus~M., 84423D, \dodoi{10.1117/12.925230}

\bibitem[{{Pontoppidan} {et~al.}(2016){Pontoppidan}, {Pickering}, {Laidler}, {Gilbert}, {Sontag}, {Slocum}, {Sienkiewicz}, {Hanley}, {Earl}, {Pueyo}, {Ravindranath}, {Karakla}, {Robberto}, {Noriega-Crespo}, \& {Barker}}]{Pontoppidan_2016}
{Pontoppidan}, K.~M., {Pickering}, T.~E., {Laidler}, V.~G., {et~al.} 2016, in Society of Photo-Optical Instrumentation Engineers (SPIE) Conference Series, Vol. 9910, Observatory Operations: Strategies, Processes, and Systems VI, ed. A.~B. {Peck}, R.~L. {Seaman}, \& C.~R. {Benn}, 991016, \dodoi{10.1117/12.2231768}

\bibitem[{{Radica} {et~al.}(2024){Radica}, {Piaulet-Ghorayeb}, {Taylor}, {Coulombe}, {Albert}, {Artigau}, {Benneke}, {Cowan}, {Doyon}, {Lafreni{\`e}re}, {L'Heureux}, \& {Lim}}]{Radica_2024}
{Radica}, M., {Piaulet-Ghorayeb}, C., {Taylor}, J., {et~al.} 2024, arXiv e-prints, arXiv:2409.19333, \dodoi{10.48550/arXiv.2409.19333}

\bibitem[{{Rudnick} \& {Gao}(2003)}]{Rudnick_2003}
{Rudnick}, R.~L., \& {Gao}, S. 2003, Treatise on Geochemistry, 3, 659, \dodoi{10.1016/B0-08-043751-6/03016-4}

\bibitem[{{Sahu} {et~al.}(2006){Sahu}, {Casertano}, {Bond}, {Valenti}, {Ed Smith}, {Minniti}, {Zoccali}, {Livio}, {Panagia}, {Piskunov}, {Brown}, {Brown}, {Renzini}, {Rich}, {Clarkson}, \& {Lubow}}]{Sahu_2006}
{Sahu}, K.~C., {Casertano}, S., {Bond}, H.~E., {et~al.} 2006, \nat, 443, 534, \dodoi{10.1038/nature05158}

\bibitem[{{Schaefer} {et~al.}(2016){Schaefer}, {Wordsworth}, {Berta-Thompson}, \& {Sasselov}}]{Schaefer_2016}
{Schaefer}, L., {Wordsworth}, R.~D., {Berta-Thompson}, Z., \& {Sasselov}, D. 2016, \apj, 829, 63, \dodoi{10.3847/0004-637X/829/2/63}

\bibitem[{Speagle(2020)}]{dynesty}
Speagle, J.~S. 2020, Monthly Notices of the Royal Astronomical Society, 493, 3132–3158, \dodoi{10.1093/mnras/staa278}

\bibitem[{{Trotta}(2008)}]{Trotta:2008}
{Trotta}, R. 2008, Contemporary Physics, 49, 71, \dodoi{10.1080/00107510802066753}

\bibitem[{{Van Looveren} {et~al.}(2024){Van Looveren}, {G{\"u}del}, {Boro Saikia}, \& {Kislyakova}}]{VanLooveren_2024}
{Van Looveren}, G., {G{\"u}del}, M., {Boro Saikia}, S., \& {Kislyakova}, K. 2024, \aap, 683, A153, \dodoi{10.1051/0004-6361/202348079}

\bibitem[{Virtanen {et~al.}(2020)Virtanen, Gommers, Oliphant, Haberland, Reddy, Cournapeau, Burovski, Peterson, Weckesser, Bright, {van der Walt}, Brett, Wilson, Millman, Mayorov, Nelson, Jones, Kern, Larson, Carey, Polat, Feng, Moore, {VanderPlas}, Laxalde, Perktold, Cimrman, Henriksen, Quintero, Harris, Archibald, Ribeiro, Pedregosa, {van Mulbregt}, \& {SciPy 1.0 Contributors}}]{2020SciPy-NMeth}
Virtanen, P., Gommers, R., Oliphant, T.~E., {et~al.} 2020, Nature Methods, 17, 261, \dodoi{10.1038/s41592-019-0686-2}

\bibitem[{{Wachiraphan} {et~al.}(2024){Wachiraphan}, {Berta-Thompson}, {Diamond-Lowe}, {Winters}, {Murray}, {Zhang}, {Xue}, {Morley}, {Rosario-Franco}, \& {Duvvuri}}]{Wachiraphan_2024}
{Wachiraphan}, P., {Berta-Thompson}, Z.~K., {Diamond-Lowe}, H., {et~al.} 2024, arXiv e-prints, arXiv:2410.10987, \dodoi{10.48550/arXiv.2410.10987}

\bibitem[{Waskom(2021)}]{seaborn}
Waskom, M.~L. 2021, Journal of Open Source Software, 6, 3021, \dodoi{10.21105/joss.03021}

\bibitem[{{Weiner Mansfield} {et~al.}(2024){Weiner Mansfield}, {Xue}, {Zhang}, {Mahajan}, {Ih}, {Koll}, {Bean}, {Coy}, {Eastman}, {Kempton}, \& {Kite}}]{WeinerMansfield_2024}
{Weiner Mansfield}, M., {Xue}, Q., {Zhang}, M., {et~al.} 2024, \apjl, 975, L22, \dodoi{10.3847/2041-8213/ad8161}

\bibitem[{{Wheatley} {et~al.}(2017){Wheatley}, {Louden}, {Bourrier}, {Ehrenreich}, \& {Gillon}}]{Wheatley_2017}
{Wheatley}, P.~J., {Louden}, T., {Bourrier}, V., {Ehrenreich}, D., \& {Gillon}, M. 2017, \mnras, 465, L74, \dodoi{10.1093/mnrasl/slw192}

\bibitem[{{Whittaker} {et~al.}(2022){Whittaker}, {Malik}, {Ih}, {Kempton}, {Mansfield}, {Bean}, {Kite}, {Koll}, {Cronin}, \& {Hu}}]{Whittaker_2022}
{Whittaker}, E.~A., {Malik}, M., {Ih}, J., {et~al.} 2022, \aj, 164, 258, \dodoi{10.3847/1538-3881/ac9ab3}

\bibitem[{{Wilson} {et~al.}(2021){Wilson}, {Froning}, {Duvvuri}, {France}, {Youngblood}, {Schneider}, {Berta-Thompson}, {Brown}, {Buccino}, {Hawley}, {Irwin}, {Kaltenegger}, {Kowalski}, {Linsky}, {Loyd}, {Miguel}, {Pineda}, {Redfield}, {Roberge}, {Rugheimer}, {Tian}, \& {Vieytes}}]{Wilson_2021}
{Wilson}, D.~J., {Froning}, C.~S., {Duvvuri}, G.~M., {et~al.} 2021, \apj, 911, 18, \dodoi{10.3847/1538-4357/abe771}

\bibitem[{{Xue} {et~al.}(2024){Xue}, {Bean}, {Zhang}, {Mahajan}, {Ih}, {Eastman}, {Lunine}, {Mansfield}, {Coy}, {Kempton}, {Koll}, \& {Kite}}]{Xue_2024}
{Xue}, Q., {Bean}, J.~L., {Zhang}, M., {et~al.} 2024, \apjl, 973, L8, \dodoi{10.3847/2041-8213/ad72e9}

\bibitem[{{Zhang} {et~al.}(2024){Zhang}, {Hu}, {Inglis}, {Dai}, {Bean}, {Knutson}, {Lam}, {Goffo}, \& {Gandolfi}}]{Zhang_2024}
{Zhang}, M., {Hu}, R., {Inglis}, J., {et~al.} 2024, \apjl, 961, L44, \dodoi{10.3847/2041-8213/ad1a07}

\bibitem[{{Zieba} {et~al.}(2023){Zieba}, {Kreidberg}, {Ducrot}, {Gillon}, {Morley}, {Schaefer}, {Tamburo}, {Koll}, {Lyu}, {Acu{\~n}a}, {Agol}, {Iyer}, {Hu}, {Lincowski}, {Meadows}, {Selsis}, {Bolmont}, {Mandell}, \& {Suissa}}]{zieba_2023}
{Zieba}, S., {Kreidberg}, L., {Ducrot}, E., {et~al.} 2023, \nat, 620, 746, \dodoi{10.1038/s41586-023-06232-z}

\end{thebibliography}
\bibliographystyle{aasjournal}



\end{document}